\documentclass[reqno,10pt,a4paper,dvips]{amsart}

\usepackage{amssymb,mathptmx,cite,eucal,array,setspace,geometry,enumitem,color}
\usepackage[dvips]{graphicx}
\usepackage{tikz}

\numberwithin{equation}{section}

\allowdisplaybreaks

\DeclareSymbolFont{largesymbols}{OMX}{zplm}{m}{n} 

\newcommand{\DSLA}[2]{\ext{\alg{#1} \left( #2 \right)}}

\newcommand{\beq}{\begin{equation}}
\newcommand{\eeq}{\end{equation}}

\newcommand{\alg}[1]{\mathfrak{#1}}
\newcommand{\group}[1]{\mathsf{#1}}

\newcommand{\func}[2]{#1 \left( #2 \right)}
\newcommand{\tfunc}[2]{#1 \bigl( #2 \bigr)}

\newcommand{\brac}[1]{\left( #1 \right)}
\newcommand{\tbrac}[1]{\bigl( #1 \bigr)}
\newcommand{\sqbrac}[1]{\left[ #1 \right]}
\newcommand{\tsqbrac}[1]{\bigl[ #1 \bigr]}
\newcommand{\set}[1]{\left\{ #1 \right\}}
\newcommand{\st}{\mspace{5mu} : \mspace{5mu}}

\newcommand{\abs}[1]{\left\lvert #1 \right\rvert}
\newcommand{\tabs}[1]{\bigl\lvert #1 \bigr\rvert}

\newcommand{\ZZ}{\mathbb{Z}}

\newcommand{\RR}{\mathbb{R}}
\newcommand{\CC}{\mathbb{C}}

\newcommand{\dd}{\mathrm{d}}
\newcommand{\ii}{\mathfrak{i}}
\newcommand{\ee}{\mathsf{e}}

\newcommand{\dCox}{\mathsf{h}^{\vee}}

\newcommand{\killing}[2]{\kappa \bigl( #1 , #2 \bigr)}
\newcommand{\invkilling}[2]{\kappa^{-1} \bigl( #1 , #2 \bigr)}

\newcommand{\affine}[1]{\widehat{#1}}

\newcommand{\comm}[2]{\bigl[ #1 , #2 \bigr]}
\newcommand{\acomm}[2]{\bigl\{ #1 , #2 \bigr\}}

\newcommand{\stcst}[2]{f^{#1}_{\hphantom{#1} #2}}

\newcommand{\ket}[1]{\bigl\lvert #1 \bigr\rangle}

\newcommand{\normord}[1]{{} : #1 : {}} 

\newcommand{\TypMod}[2]{\overline{\mathcal{T}}^{#1}_{#2}}
\newcommand{\StypMod}[2]{\overline{\mathcal{S}}^{#1}_{#2}}
\newcommand{\AtypMod}[2]{\overline{\mathcal{A}}^{#1}_{#2}}
\newcommand{\GenTypMod}[3]{{}^{\brac{#3}}\overline{\mathcal{T}}^{#1}_{#2}}
\newcommand{\VerMod}[2]{\overline{\mathcal{V}}^{#1}_{#2}}

\newcommand{\ProjMod}[2]{\overline{\mathcal{P}}^{#1}_{#2}}

\newcommand{\AffTypMod}[2]{\mathcal{T}^{#1}_{#2}}
\newcommand{\AffStypMod}[2]{\mathcal{S}^{#1}_{#2}}
\newcommand{\AffAtypMod}[2]{\mathcal{A}^{#1}_{#2}}
\newcommand{\AffGenTypMod}[3]{{}^{\brac{#3}}\mathcal{T}^{#1}_{#2}}
\newcommand{\AffVerMod}[2]{\mathcal{V}^{#1}_{#2}}
\newcommand{\AffProjMod}[2]{\mathcal{P}^{#1}_{#2}}

\newcommand{\SLA}[2]{\alg{#1} \left( #2 \right)}
\newcommand{\SLG}[2]{\group{#1} \left( #2 \right)}
\newcommand{\SLSA}[3]{\alg{#1} \left( #2 \middle\vert #3 \right)}

\newcommand{\DSLSA}[3]{\ext{\alg{#1} \left( #2 \middle\vert #3 \right)}}
\newcommand{\AKMA}[2]{\affine{\alg{#1}} \left( #2 \right)}
\newcommand{\AKMSA}[3]{\affine{\alg{#1}} \left( #2 \middle\vert #3 \right)}
\newcommand{\DAKMA}[2]{\ext{\alg{#1}} \left( #2 \right)}
\newcommand{\DAKMSA}[3]{\ext{\alg{#1}} \left( #2 \middle\vert #3 \right)}

\newcommand{\traceover}[1]{\tr_{\raisebox{-3pt}{$\scriptstyle #1$}}}

\newcommand{\ch}[1]{\mathrm{ch} \bigl[ #1 \bigr]}
\newcommand{\fch}[2]{\ch{#1} \bigl( #2 \bigr)}
\newcommand{\sch}[1]{\mathrm{sch} \bigl[ #1 \bigr]}

\newcommand{\jth}[1]{\vartheta_{#1}}
\newcommand{\Jth}[2]{\jth{#1} \bigl( #2 \bigr)}

\newcommand{\modS}{\mathsf{S}}
\newcommand{\modT}{\mathsf{T}}

\newcommand{\SMat}[4]{\modS \left[ \begin{smallmatrix} #1 \vphantom{#3} \\ #2 \vphantom{#4} \end{smallmatrix} \: ; \: \begin{smallmatrix} #3 \vphantom{#1} \\ #4 \vphantom{#2} \end{smallmatrix} \right]}
\newcommand{\SMatStyp}[4]{\modS \left[ \overline{\begin{smallmatrix} #1 \vphantom{#3} \\ #2 \vphantom{#4} \end{smallmatrix}} \: ; \: \begin{smallmatrix} #3 \vphantom{#1} \\ #4 \vphantom{#2} \end{smallmatrix} \right]}
\newcommand{\SMatAtyp}[4]{\modS \left[ \overline{\overline{\begin{smallmatrix} #1 \vphantom{#3} \\ #2 \vphantom{#4} \end{smallmatrix}}} \: ; \: \begin{smallmatrix} #3 \vphantom{#1} \\ #4 \vphantom{#2} \end{smallmatrix} \right]}

\newcommand{\modarg}[7]{\left( \: #1 \: \middle\vert \: #2 \: \middle\vert \: #3 \: %
                  \middle\Vert \: #4 \: \middle\vert \: #5 \: \middle\vert \: #6 \: %
                  \middle\Vert \: #7 \: \right)}

\newcommand{\fuse}{\mathbin{\times_{\! f}}}
\newcommand{\fuscoeff}[2]{\mathsf{N}_{#1}^{\hphantom{#1} #2}}

\newcommand{\ses}[3]{0 \longrightarrow #1 \longrightarrow #2 \longrightarrow #3 \longrightarrow 0}
\newcommand{\ftes}[4]{0 \longrightarrow #1 \longrightarrow #2 \longrightarrow #3 \longrightarrow #4 \longrightarrow 0}
\newcommand{\res}[5]{\cdots \longrightarrow #5 \longrightarrow #4 \longrightarrow #3 \longrightarrow #2 \longrightarrow #1 \longrightarrow 0}

\newcommand{\eqnref}[1]{Equation~\eqref{#1}}
\newcommand{\eqnDref}[2]{Equations~\eqref{#1} and \eqref{#2}}

\newcommand{\secref}[1]{Section~\ref{#1}}

\newcommand{\figref}[1]{Figure~\ref{#1}}

\newcommand{\exref}[1]{Example~\ref{#1}}

\newcommand{\propref}[1]{Proposition~\ref{#1}}

\newcommand{\thmref}[1]{Theorem~\ref{#1}}

\newcommand{\cft}{conformal field theory}
\newcommand{\cfts}{conformal field theories}
\newcommand{\uea}{universal enveloping algebra}
\newcommand{\ueas}{universal enveloping algebras}
\newcommand{\lcft}{logarithmic conformal field theory}
\newcommand{\lcfts}{logarithmic conformal field theories}

\newcommand{\hws}{highest weight state}
\newcommand{\hwss}{highest weight states}
\newcommand{\lws}{lowest weight state}
\newcommand{\lwss}{lowest weight states}

\newcommand{\hwms}{highest weight modules}

\newcommand{\ext}[1]{\widetilde{#1}}
\newcommand{\tE}{\ext{E}}
\newcommand{\tF}{\ext{F}}
\newcommand{\tH}{\ext{H}}
\newcommand{\tJ}{\ext{J}}
\newcommand{\tK}{\ext{K}}
\newcommand{\tN}{\ext{N}}
\newcommand{\ta}{\ext{a}}
\newcommand{\te}{\ext{e}}
\newcommand{\tk}{\ext{k}}
\newcommand{\tn}{\ext{n}}
\newcommand{\ttt}{\ext{t}}
\newcommand{\tx}{\ext{x}}
\newcommand{\ty}{\ext{y}}
\newcommand{\tz}{\ext{z}}
\newcommand{\tmu}{\ext{\mu}}
\newcommand{\tnu}{\ext{\nu}}
\newcommand{\tsigma}{\ext{\sigma}}
\newcommand{\tpsi}{\ext{\psi}}

\newcommand{\sfmod}[2]{\tfunc{\tsigma^{#1}}{#2}}

\DeclareMathOperator{\id}{id}
\DeclareMathOperator{\vectspan}{span}
\DeclareMathOperator{\tr}{tr}

\DeclareMathOperator{\sdim}{sdim}

\renewcommand{\Re}{\operatorname{Re}}

\theoremstyle{plain}
\newtheorem{ex}{Example}
\newtheorem{thm}{Theorem}
\newtheorem{prop}[thm]{Proposition}
\newtheorem{cor}[thm]{Corollary}
\newtheorem{lem}[thm]{Lemma}

\begin{document}

\title{Takiff Superalgebras and Conformal Field Theory}

\author[A Babichenko]{Andrei Babichenko}

\address[Andrei Babichenko]{
Department of Particle Physics, Weizmann Institute \\
Rehovot 76100, Israel \\
and
Department of Mathematics, University of York \\
Heslington, York, YO10 5DD, UK
}

\email{babichen@weizmann.ac.il}

\author[D Ridout]{David Ridout}

\address[David Ridout]{
Department of Theoretical Physics \\
Research School of Physics and Engineering \\
and
Department of Mathematics \\
Mathematical Sciences Institute \\
Australian National University \\
Canberra, ACT 0200 \\
Australia
}

\email{david.ridout@anu.edu.au}

\thanks{\today}

\begin{abstract}
A class of non-semisimple extensions of Lie superalgebras is studied.  They are obtained by adjoining to the superalgebra its adjoint representation as an abelian ideal.  When the superalgebra is of affine Kac-Moody type, a generalisation of Sugawara's construction is shown to give rise to a copy of the Virasoro algebra and so, presumably, to a conformal field theory.  Evidence for this is detailed for the extension of the affinisation of the superalgebra $\SLSA{gl}{1}{1}$:  Its highest weight irreducible modules are classified using spectral flow, the irreducible supercharacters are computed and a continuum version of the Verlinde formula is verified to give non-negative integer structure coefficients.  Interpreting these coefficients as those of the Grothendieck ring of fusion, partial results on the true fusion ring and its indecomposable structures are deduced.
\end{abstract}

\maketitle

\onehalfspacing

\section{Introduction}

Recent investigations have shown that logarithmic conformal field theories are playing an essential role in different physical problems ranging from string theory, especially on supergroup target spaces \cite{RozSTM93,SalGL106,GotWZN07,QueFre07,MQS08,CMQSS09,MQS11}, to different condensed matter and statistical mechanics problems \cite{Re01,PirBou05,PeaLog06,ReaAss07,RidPer07,GaiCon11,VJS12,GJSV12,GRS12}.  On the one hand, the increasing interest in such theories is not surprising given that they are expected to be generic.  On the other, recent progress in their understanding, mainly achieved by more careful studies of the representation theory of their (chiral) symmetry algebras \cite{RohRed96,AdaTri08,RidSta09,GS11}, has added considerably to their appeal.

The defining feature of a logarithmic theory is the presence of reducible but indecomposable representations on which the Virasoro zero mode $L_0$ acts non-semisimply, leading to logarithmic singularities in correlation functions \cite{GurLog93}.  In this paper, we introduce a new class of non-semisimple symmetry algebras which we expect to lead to new examples of reasonably well-behaved \lcfts{} and we confirm this expectation for one non-trivial example.  This class consists of certain types of extensions of Kac-Moody superalgebras in which, roughly speaking, the superalgebra is extended by its adjoint representation.

We remark that similar extensions have recently been proposed \cite{Ash09,Ben10} as part of a current algebra description of two-dimensional principal chiral models for Lie supergroups with vanishing Killing form (see also \cite{Kon10,CanAno12}).  These deformations of Wess-Zumino-Witten models are known to be conformal \cite{Ber99,BerZhu99,Ba06} due to their beta function vanishing perturbatively to all orders.  However, the currents postulated in this proposal are \emph{non-chiral}, whereas we build our theories from a traditional chiral starting point (we also choose a more useful energy-momentum tensor to define the conformal structure).  It is therefore not clear if our results will have any bearing on the understanding of the \cfts{} describing these principal chiral models.

In the mathematics literature, the type of non-semisimple Lie superalgebras which arise in the above investigations were introduced by Takiff \cite{Tak71}, though not in the super setting, as part of an investigation of invariant polynomial rings.  These algebras have since been considered in a slightly generalised form under the names \emph{generalised Takiff algebras} \cite{Rais,Geoffriau}, in which a semisimple Lie algebra is tensored with a polynomial ring in a nilpotent formal variable $t$, and \emph{truncated current algebras} \cite{Wil07,Wil08}, in which one does the same to an affine Kac-Moody algebra.  The algebras that we will consider correspond to taking $t^2=0$, as in Takiff's original paper, but we extend to embrace Lie and Kac-Moody superalgebras as well.  We will therefore refer to them as \emph{Takiff superalgebras}.  Unfortunately, not much is known about the representations of Takiff algebras beyond the highest weight theory \cite{Wil07}.

With this motivation, we provide the first steps in investigating the representation theory of Takiff superalgebras as required by the intended application to conformal field theory.  For the most part, we restrict ourselves to studying the Takiff superalgebras obtained from the non-semisimple Lie superalgebra $\SLSA{gl}{1}{1}$ and its affinisation $\AKMSA{gl}{1}{1}$.  The logarithmic conformal field theories built from $\AKMSA{gl}{1}{1}$ are among the best understood \cite{RozSTM93,SalGL106,GotRep07,CreBra08,CreGL109} and the representations may be classified using relatively elementary means \cite{CreRel11} (see also \cite{KacSup87} for a somewhat less elementary discussion).  Our reason for not immediately generalising to the Takiff algebra of $\AKMA{sl}{2}$ or the Takiff superalgebra of $\AKMSA{psl}{2}{2}$, for example, is the realisation \cite{FeiEqu98,MalStr01,GabFus01,RidSL210,CreMod12} that logarithmic conformal field theories built from affine Kac-Moody algebras (and superalgebras) generically require the introduction of irreducible representations that are not highest weight with respect to \emph{any} choice of Borel subalgebra.\footnote{That $\AKMSA{gl}{1}{1}$ provides an exception to this expectation may be understood as resulting from the special feature of $\SLSA{gl}{1}{1}$ that the raising and lowering operators are all nilpotent.}  We expect that this will be true for Takiff superalgebras as well and therefore leave the difficult problem of characterising such representations for future work \cite{BR}.

The structure of the article is as follows:  We first define a Takiff superalgebra precisely and then show, for a reasonably general class of affine Takiff superalgebras, that there is a natural generalisation of the Sugawara construction which gives an energy-momentum tensor.  Moreover, this field has the desired property that each current is a dimension $1$ primary field with respect to it.  Whether this energy-momentum field is physically relevant or not in any given application, the construction is extremely useful for the investigation of the representation theory that follows.  We then specialise to the Takiff superalgebra constructed from $\SLSA{gl}{1}{1}$.  Keeping in mind potential physical applications, we analyse the structure of its Verma modules and classify the irreducible highest weight modules.  This is followed by a discussion of the representation ring generated by repeatedly taking tensor products of irreducibles.  In contrast to the case of $\SLSA{gl}{1}{1}$ \cite{SalGL106,GotRep07}, the complexity of the indecomposable modules that appear as summands of such tensor products appears to grow without limit (more precisely, the action of one of the affine zero modes appears to involve Jordan blocks of ever-increasing rank).  Nevertheless, we can completely characterise the \emph{Grothendieck ring} associated to the representation ring.

We next turn to the Takiff superalgebra of the affine Kac-Moody superalgebra $\AKMSA{gl}{1}{1}$ and a classification of its irreducible highest weight modules.  The main problem here is to understand the submodule structure of the Verma modules and this follows readily from a study of the effect of twisting representations by certain ``spectral flow'' automorphisms (we follow \cite{SalGL106,CreRel11}).  As one expects, the Verma modules are generically irreducible leading to a notion of \emph{typicality} for irreducibles.  Unlike the non-Takiff $\AKMSA{gl}{1}{1}$ case, here we find two distinct non-trivial submodule structures leading to irreducibles that we christen \emph{semitypical} and \emph{atypical}.  They correspond, in superalgebra language, to modules of atypicality degree $1$ and $2$, respectively.  Based on these structures, we compute exact sequences realising the irreducibles in terms of Verma modules (Bern\v{s}te\u{\i}n-Gel'fand-Gel'fand resolutions) and so obtain character formulae.

One of the central questions in any formal construction of a conformal field theory is that of the modular transformation properties of the characters.  The full spectrum of representations of an affine Takiff superalgebra will be much larger than just the irreducibles, even in the case of $\AKMSA{gl}{1}{1}$.  In particular, fusion will generate a veritable zoo of indecomposable modules of different natures from the irreducibles.  Given that we did not exhaustively list the indecomposables of the Takiff superalgebra of $\SLSA{gl}{1}{1}$, it is not reasonable to expect such a list in the affine case.  However, characters do not distinguish between direct and indecomposable sums, so it is reasonable to ask after the \emph{Grothendieck ring} associated to the fusion ring.  Here, one expects that there will be some sort of Verlinde-type formula with which one can compute the structure constants of this Grothendieck ring.  We find the S- and T-matrices describing the modular transformations of the affine Takiff superalgebra supercharacters and show that the obvious generalisation of the Verlinde formula results in non-negative integer structure constants.  We therefore conjecture that these constants are those of the Grothendieck fusion ring.  This work concludes with a brief discussion of what this means for the genuine fusion ring and how fusion results can be verified, before describing our conclusions and speculating on future work.

\section{Takiff Superalgebras}

Let $\alg{g}$ be a Lie algebra.  We adjoin the \emph{adjoint representation} of $\alg{g}$ to the algebra itself, making the result into a Lie algebra by declaring that the bracket of any two elements of the adjoint representation is zero.  More explicitly, if $\set{J^a}$ denotes a basis of $\alg{g}$ with structure constants $\stcst{ab}{c}$,
\begin{equation}
\comm{J^a}{J^b} = \sum_c \stcst{ab}{c} \: J^c,
\end{equation}
then we extend this basis by new elements $\set{\tJ^a}$, where the parity of $\tJ^a$ matches that of $J^a$, and impose
\begin{equation}
\comm{J^a}{\tJ^b} = \sum_c \stcst{ab}{c} \: \tJ^c, \qquad \comm{\tJ^a}{\tJ^b} = 0.
\end{equation}
It is easy to check that the Jacobi identity is satisfied, hence that the extended basis spans a Lie algebra.  Indeed, the result may be characterised as a semidirect sum of $\alg{g}$ with itself \cite{Jac62}.  We denote this Lie algebra by $\ext{\alg{g}}$ and will refer to it as the \emph{Takiff algebra} of $\alg{g}$.  It is clear that this construction may be extended to Lie superalgebras by changing the commutators above to graded commutators and checking the graded Jacobi identity.  We will then speak of the \emph{Takiff superalgebra} of a Lie superalgebra $\alg{g}$.

Such algebras were studied in \cite{Tak71} and were subsequently generalised as follows:
\begin{equation}
\alg{g} \left\langle m \right\rangle = \alg{g} \otimes \frac{\CC \sqbrac{t}}{t^m \CC \sqbrac{t}}, \qquad \comm{x \otimes t^i}{y \otimes t^j} = \comm{x}{y} \otimes t^{i+j}.
\end{equation}
Because of this, these generalised Takiff algebras are often referred to as polynomial Lie algebras or truncated current algebras.  We will only consider the original Takiff algebras ($m=2$) in what follows.

\begin{ex}
Let $\alg{g}$ be the one-dimensional abelian Lie algebra $\SLA{u}{1}$ with basis element $a$.  Then, the Takiff algebra $\DSLA{u}{1}$ is two-dimensional, with basis $\set{a,\ta}$ and the Lie bracket is given by
\begin{equation}
\comm{a}{\ta} = 0.
\end{equation}
Clearly, $\DSLA{u}{1}$ is isomorphic to the abelian Lie algebra $\SLA{u}{1} \oplus \SLA{u}{1}$.
\end{ex}

\noindent This example is typical for abelian Lie superalgebras --- the Takiff superalgebra is always the direct sum of two copies of the superalgebra.  Takiff superalgebras are more interesting in the non-abelian setting.

\begin{ex} \label{ex:ExtSL2}
Recall that $\SLA{sl}{2}$ has a standard basis $\set{E,H,F}$ for which the non-trivial commutation relations are
\begin{equation} \label{eqn:CommSL2}
\comm{H}{E} = 2E, \qquad \comm{E}{F} = H, \qquad \comm{H}{F} = -2F.
\end{equation}
The Takiff algebra $\DSLA{sl}{2}$ is then six-dimensional with basis $\set{E,H,F,\tE,\tH,\tF}$.  The non-trivial commutation relations are those of \eqref{eqn:CommSL2} along with
\begin{equation} \label{eqn:CommExtSL2}
\comm{H}{\tE} = \comm{\tH}{E} = 2 \tE, \qquad \comm{E}{\tF} = \comm{\tE}{F} = \tH, \qquad \comm{H}{\tF} = \comm{\tH}{F} = -2 \tF.
\end{equation}
However, it is not hard to check that $\DSLA{sl}{2}$ does not decompose as a (non-trivial) direct sum of ideals.  In particular, the span of $\set{\tE,\tH,\tF}$ is an abelian ideal of $\DSLA{sl}{2}$ with no complement (the commutant of this ideal is precisely the ideal itself).
\end{ex}

\begin{ex} \label{ex:ExtGL11}
The Lie superalgebra $\SLSA{gl}{1}{1}$ is spanned by two even (bosonic) elements $N$ and $E$ and two odd (fermionic) elements $\psi^+$ and $\psi^-$.  The non-trivial relations are
\begin{equation} \label{eqn:CommGL11}
\comm{N}{\psi^{\pm}} = \pm \psi^{\pm}, \qquad \acomm{\psi^+}{\psi^-} = E.
\end{equation}
We therefore obtain the Takiff superalgebra $\DSLSA{gl}{1}{1}$ by adjoining even elements $\tN$ and $\tE$ and odd elements $\tpsi^+$ and $\tpsi^-$, subject to \eqref{eqn:CommGL11} and
\begin{equation} \label{eqn:CommExtGL11}
\comm{N}{\tpsi^{\pm}} = \comm{\tN}{\psi^{\pm}} = \pm \tpsi^{\pm}, \qquad
\acomm{\psi^+}{\tpsi^-} = \acomm{\tpsi^+}{\psi^-} = \tE,
\end{equation}
with all other brackets vanishing.  Again, the Takiff superalgebra does not decompose as a direct sum of ideals.
\end{ex}

Our interest here lies in the Takiff superalgebras of affine Kac-Moody superalgebras.  Let us therefore suppose that $\alg{g}$ is a finite-dimensional basic classical simple complex Lie superalgebra.  This includes the finite-dimensional simple complex Lie algebras as special cases, while being basic and classical ensures that there is a non-degenerate even supersymmetric bilinear form $\killing{\cdot}{\cdot}$ in the general case (see \cite{CorGro89}).  The affinisation of $\alg{g}$ is then the Lie superalgebra
\begin{subequations} \label{eqn:DefAff}
\begin{equation}
\affine{\alg{g}} = \brac{\alg{g} \otimes \CC \sqbrac{t ; t^{-1}}} \oplus \vectspan_{\CC} \set{K}
\end{equation}
with the (graded) bracket
\begin{equation}
\comm{J^a \otimes t^m}{J^b \otimes t^n} = \sum_c \stcst{ab}{c} J^c \otimes t^{m+n} + m \killing{J^a}{J^b} \delta_{m+n,0} K, \qquad \comm{J^a \otimes t^m}{K} = 0.
\end{equation}
\end{subequations}
The central element $K$ will be assumed to act as a fixed multiple $k$ of the identity in the representations that are of interest.  This multiple $k$ is called the \emph{level}.

\begin{ex}
The affine Kac-Moody algebra $\AKMA{u}{1}$ has basis $\set{a_n = a \otimes t^n , K \st n \in \ZZ}$ and commutation relations
\begin{equation} \label{eqn:CommAffU1}
\comm{a_m}{a_n} = m \delta_{m+n,0} K.
\end{equation}
Here, we have normalised $\killing{a}{a} = 1$.  The Takiff algebra $\DAKMA{u}{1} \equiv \ext{\AKMA{u}{1}}$ then has $\set{a_n , \ta_n , K , \tK \st n \in \ZZ}$ for a basis and the non-trivial commutation relations are \eqref{eqn:CommAffU1} and
\begin{equation} \label{eqn:CommExtU1}
\comm{a_m}{\ta_n} = m \delta_{m+n,0} \tK.
\end{equation}
Because $\DAKMA{u}{1}$ is (mildly) non-abelian, it does not
decompose as a direct sum of ideals.  However, if we restrict to a
category of representations on which $K$ and $\tK$ act as fixed
multiples $k$ and $\tk$ (respectively) of the identity, then we
may check that the $a_m$ and the $b_n = k \ta_n - \tk a_n$ act on
these representations as \emph{commuting} copies of $\AKMA{u}{1}$.
We therefore conclude that when $k$ and $\tk$ are non-zero, such a
category of representations may be regarded as a category of
$\AKMA{u}{1} \oplus \AKMA{u}{1}$-representations.
\end{ex}

\noindent This observation generalises to the affinisations of other abelian Lie superalgebras.  While the Takiff superalgebra is technically indecomposable, its action on representations for which $K$ and $\tK$ act as multiples of of the identity may be replaced by an action of the direct sum of two copies of the affinisation.  As we are more interested in the representation theory of Takiff superalgebras, rather than in these superalgebras themselves, we shall turn to the analysis of non-abelian examples.  In fact, the majority of this article is devoted to the representation theory of the affinisation of the Lie superalgebra discussed in \exref{ex:ExtGL11}.

Before commencing these analyses, we pause to indicate one reason why Takiff superalgebras could be of interest to field theorists.  Recall that an important fact, both for physics and mathematics, about a Kac-Moody superalgebra $\affine{\alg{g}}$, based on a simple Lie superalgebra $\alg{g}$, is that their level $k$ \ueas{} contain a copy of the \uea{} of the Virasoro algebra with central charge
\begin{equation} \label{eqn:SugawaraC}
c = \frac{k \sdim \alg{g}}{k + \dCox},
\end{equation}
where $\dCox$ is the dual Coxeter number of $\alg{g}$ (and $k \neq -\dCox$) and $\sdim$ denotes the superdimension.  This gives rise to a vertex superalgebra structure for Kac-Moody superalgebras which underlies a vast amount of important work in pure mathematics and \cft{}.

The explicit embedding of the Virasoro algebra into the \uea{} of $\affine{\alg{g}}$ is known as the Sugawara construction and is best described field-theoretically.  If $\set{J^a}$ is any basis of $\alg{g}$, then we form fields (generating functions) in an indeterminate $z$ from the elements $J^a_n = J^a \otimes t^n \in \affine{\alg{g}}$ as follows:
\begin{equation}
\func{J^a}{z} = \sum_{n \in \ZZ} J^a_n z^{-n-1}.
\end{equation}
From these fields, one defines
\begin{equation}
\func{T}{z} = \frac{1}{2 \brac{k + \dCox}} \sum_{a,b} \invkilling{J^a}{J^b} \normord{\func{J^a}{z} \func{J^b}{z}}.
\end{equation}
Here, $\invkilling{\cdot}{\cdot}$ is the bilinear form on $\alg{g}$ which is inverse to $\killing{\cdot}{\cdot}$ and the normally-ordered product of the fields is given, at the level of the modes $J^a_m$ and $J^b_n$ ($z$ is left invariant), by
\begin{equation}
\normord{J^a_m J^b_n} =
\begin{cases}
J^a_m J^b_n & \text{if $m \leqslant -1$,} \\
\brac{-1}^{p^a p^b} J^b_n J^a_m & \text{if $m \geqslant 0$,}
\end{cases}
\end{equation}
where $p^a$ denotes the parity of $J^a$ ($p^a = 0$ if $J^a$ is even, $p^a = 1$ if $J^a$ is odd).  The standard basis elements $L_n$ of the Virasoro algebra are then, finally, recovered from
\begin{equation} \label{eqn:VirModes}
T(z) = \sum_{n \in \ZZ} L_n z^{-n-2}
\end{equation}
and the central charge is found to be that given in \eqref{eqn:SugawaraC}.

It is extremely interesting to observe that this generalises in a not-unpleasant fashion to the Takiff superalgebras of the affine Kac-Moody superalgebras.  The analogue of the Sugawara construction is as follows:

\begin{thm} \label{thm:ExtSugawara}
If $\set{J^a}$ denotes a basis of a finite-dimensional basic classical simple complex Lie superalgebra $\alg{g}$, $\set{J^a_n , K \st n \in \ZZ}$ the induced basis of the affinisation $\affine{\alg{g}}$, and $\set{J^a_n , \tJ^a_n , K , \tK \st n \in \ZZ}$ the induced basis of the Takiff superalgebra $\ext{\affine{\alg{g}}}$ of the affinisation, then the field
\begin{equation} \label{eqn:DoubledT}
\func{T}{z} = \frac{1}{\tk} \sum_{a,b} \invkilling{J^a}{J^b} \normord{\func{J^a}{z} \func{\tJ^b}{z}} - \frac{k + 2 \dCox}{2 \tk^2} \sum_{a,b} \invkilling{J^a}{J^b} \normord{\func{\tJ^a}{z} \func{\tJ^b}{z}}
\end{equation}
has modes $L_n$ as in \eqref{eqn:VirModes} satisfying the Virasoro commutation relations with central charge
\begin{equation} \label{eqn:DoubledC}
c = 2 \sdim \alg{g}.
\end{equation}
Moreover,
\begin{equation}
\comm{L_m}{J^a_n} = -n J^a_{m+n}, \qquad \comm{L_m}{\tJ^a_n} = -n \tJ^a_{m+n}.
\end{equation}
\end{thm}

\noindent The proof is an easy extension of the usual proof of the Sugawara construction (see \cite{DifCon97} for example) and we expect that an analogue for generalised affine Takiff superalgebras should also exist.  Because of this result, one expects to be able to construct \cfts{} with these Takiff superalgebras as chiral algebras.  We remark that we have checked for a few low-rank simple $\alg{g}$ that the field $\func{T}{z}$ given above is the \emph{unique} field with these properties.

\section{The Takiff Superalgebra of $\SLSA{gl}{1}{1}$} \label{sec:ExtFinGL11}

In this section, we study the representation theory of the Takiff superalgebra $\DSLSA{gl}{1}{1}$ introduced in \exref{ex:ExtGL11}.  This will be followed by a detailed study of the Takiff superalgebra of the affine Kac-Moody algebra $\AKMSA{gl}{1}{1}$ in \secref{sec:ExtGL11}.

\subsection{Irreducible Representations of $\DSLSA{gl}{1}{1}$} \label{sec:FinGL11Reps}

Recall from \exref{ex:ExtGL11} that the Takiff superalgebra $\DSLSA{gl}{1}{1}$ is spanned by four bosonic elements, $N$, $\tN$, $E$ and $\tE$, and four fermionic elements, $\psi^+$, $\tpsi^+$, $\psi^-$ and $\tpsi^-$, subject to the relations given in \eqnDref{eqn:CommGL11}{eqn:CommExtGL11}.  There is an obvious triangular decomposition:
\begin{equation} \label{eqn:FinGL11TriDec}
\DSLSA{gl}{1}{1} = \vectspan \set{\psi^- , \tpsi^-} \oplus \vectspan \set{N,E,\tN,\tE} \oplus \vectspan \set{\psi^+ , \tpsi^+} \qquad \text{(as subalgebras)}.
\end{equation}
Thus, we regard $\psi^+$ and $\tpsi^+$ as raising operators, $\psi^-$ and $\tpsi^-$ as lowering operators, and $N$, $E$, $\tN$ and $\tE$ as generating the Cartan subalgebra.  It is clear that $E$ and $\tE$ are central.  Modulo polynomials in the central generators, there are two linearly independent quadratic Casimirs in the \uea{} which we may take to be
\begin{equation} \label{eqn:Casimirs}
Q_1 = N \tE + \tN E + \psi^- \tpsi^+ + \tpsi^- \psi^+, \qquad Q_2 = \tN \tE + \tpsi^- \tpsi^+.
\end{equation}
It is important to note that while the basis we have chosen diagonalises the adjoint action of $N$, $E$ and $\tE$, it does not diagonalise that of $\tN$:
\begin{equation}
\comm{\tN}{\psi^{\pm}} = \pm \tpsi^{\pm}.
\end{equation}
It follows that $\tN$ will act non-semisimply on the adjoint module.  Moreover, if we split $\tN$ into a semisimple and a nilpotent part, the former is seen to be central.

This triangular decomposition allows us to define \hwss{} in the usual manner.  We will define a \hws{} to be an eigenvector of $N$, $E$, $\tN$ and $\tE$ which is annihilated by both raising operators $\psi^+$ and $\tpsi^+$.  The eigenvalues of the action of the Cartan basis elements are denoted by $n$, $e$, $\tn$ and $\te$, respectively.  A \lws{} is, similarly, an eigenvector of $N$, $E$, $\tN$ and $\tE$ which is annihilated by the lowering operators $\psi^-$ and $\tpsi^-$.  A \hws{} then generates a highest weight Verma module $\VerMod{n,e}{\tn,\te}$ through the free action of the lowering operators.  One similarly obtains lowest weight Verma modules by freely acting with the raising operators on a \lws{}.  We shall mostly concern ourselves with highest weight Verma modules in what follows, understanding that this is what is meant when ``highest weight'' is omitted.

Since the lowering operators $\psi^-$ and $\tpsi^-$ both square to zero and anticommute with one another, it follows that every Verma module is four-dimensional.  We illustrate the generation of a Verma module from a \hws{} $\ket{v}$ as follows:
\begin{equation} \label{pic:Verma}
\parbox[c]{0.3\textwidth}{
\begin{center}
\begin{tikzpicture}[auto,thick]
\node (top) at (0,1.5) [] {$\ket{v}$};
\node (left) at (-1.5,0) [] {$\psi^- \ket{v}$};
\node (right) at (1.5,0) [] {$\tpsi^- \ket{v}$};
\node (bot) at (0,-1.5) [] {$\tpsi^- \psi^- \ket{v}$};
\draw [->] (top) to (left);
\draw [->] (top) to (right);
\draw [->] (left) to (bot);
\draw [->] (right) to (bot);
\draw [->,dotted] (left) to (right);
\end{tikzpicture}
\end{center}
}
.
\end{equation}
All of the basis states are eigenvalues of $\tN$ except for $\psi^- \ket{v}$ (hence the dotted arrow):
\begin{equation}
\tN \psi^- \ket{v} = \tn \psi^- \ket{v} - \tpsi^- \ket{v}.
\end{equation}
Here, $\tn$ denotes the eigenvalue of $\tN$ on $\ket{v}$.  Despite this non-semisimple action, the Casimirs $Q_1$ and $Q_2$ act on this Verma module as multiplication by $n \te + \tn e$ and $\tn \te$, respectively.

We remark that we could have generalised the above notion of \hws{} and Verma module by relaxing the condition that the former must be an eigenstate of $\tN$.  However, if the resulting generalised Verma module is to be finite-dimensional, the Jordan cell for $\tN$ involving the generalised \hws{} will contain a genuine \hws{}, showing that a standard Verma module will appear as a submodule.  It follows that a finite-dimensional generalised Verma module is always realised as an indecomposable sum of standard Verma modules.  In particular, one will obtain no irreducible quotients from a generalised Verma module that could not already be obtained from a standard Verma module.  As our first aim is to understand these irreducibles, this justifies the definition of \hwss{} that we have given above.

\begin{prop} \label{prop:FinGL11Irreps}
Consider the Verma module $\VerMod{n,e}{\tn,\te}$ generated from a \hws{} $\ket{v}$.  There are three possibilities for the irreducible quotient of this Verma module:
\begin{itemize}[leftmargin=*]
\item If $e = \te = 0$, then $\tpsi^- \ket{v}$ is a singular vector and $\psi^- \ket{v}$ is a generalised singular vector, meaning that it only fails to be an eigenvector for $\tN$.  The latter vector generates the maximal submodule, and the irreducible quotient is $1$-dimensional.  We will denote this irreducible by $\AtypMod{n,0}{\tn,0}$ and refer to it as being \emph{atypical}.
\item If $\te = 0$ but $e \neq 0$, then $\tpsi^- \ket{v}$ is the only singular vector, up to scalar multiples, and it generates the maximal submodule.  The irreducible quotient is then $2$-dimensional and we will denote it by $\StypMod{n-1/2,e}{\tn,0}$ and refer to it as being \emph{semitypical}.
\item If $\te \neq 0$, then there are no (non-trivial) singular vectors, hence the irreducible quotient is $4$-dimensional:  The Verma module is itself irreducible.  We will denote this irreducible by $\TypMod{n-1,e}{\tn,\te}$ and refer to it as being \emph{typical}.
\end{itemize}
\end{prop}

\noindent We remark that the case in which $e=0$ but $\te \neq 0$ is \emph{typical} (because the Verma module has no non-trivial singular vectors).  We further remark that $\tN$ is diagonalisable on the non-typical irreducibles.  Finally, the labelling of the $N$-eigenvalue that we have adopted above deserves some comment.  It turns out to be convenient in the long run to use the \emph{average} of the $N$-eigenvalues of the basis vectors rather than that of some particular generator (such as the \hws{}).

\subsection{Tensor Products of $\DSLSA{gl}{1}{1}$-Modules} \label{sec:TPFinGL11Reps}

We are interested in the representation ring generated by these irreducibles under the graded tensor product.  It is clear that the tensor products involving the one-dimensional atypical irreducibles are rather trivial to calculate:
\begin{equation} \label{TP:Ax}
\AtypMod{n_1,0}{\tn_1,0} \otimes \AtypMod{n_2,0}{\tn_2,0} = \AtypMod{n_1+n_2,0}{\tn_1+\tn_2,0}, \qquad
\AtypMod{n_1,0}{\tn_1,0} \otimes \StypMod{n_2,e_2}{\tn_2,0} = \StypMod{n_1+n_2,e_2}{\tn_1+\tn_2,0}, \qquad
\AtypMod{n_1,0}{\tn_1,0} \otimes \TypMod{n_2,e_2}{\tn_2,\te_2} = \TypMod{n_1+n_2,e_2}{\tn_1+\tn_2,\te_2}.
\end{equation}
The first non-trivial tensor product is therefore $\StypMod{n_1,e_1}{\tn_1,0} \otimes \StypMod{n_2,e_2}{\tn_2,0}$, whose decomposition depends upon the typicality of $e_1+e_2$.  To be more specific, let $\ket{v}$ and $\ket{w}$ be the \hwss{} of $\StypMod{n_1,e_1}{\tn_1,0}$ and $\StypMod{n_2,e_2}{\tn_2,0}$, respectively.  Then for $e_1+e_2 \neq 0$, there are two unrelated \hwss{}:
\begin{equation}
\ket{v} \otimes \ket{w} \qquad \text{and} \qquad
e_1 \ket{v} \otimes \psi^- \ket{w} - e_2 \psi^- \ket{v} \otimes \ket{w}.
\end{equation}
When $e_1+e_2 = 0$ however, the latter becomes a singular descendant of the former.  The tensor product decomposition is then
\begin{equation} \label{TP:SxS}
\StypMod{n_1,e_1}{\tn_1,0} \otimes \StypMod{n_2,e_2}{\tn_2,0} =
\begin{cases}
\StypMod{n_1+n_2+1/2,e_1+e_2}{\tn_1+\tn_2,0} \oplus \StypMod{n_1+n_2-1/2,e_1+e_2}{\tn_1+\tn_2,0} & \text{if $e_1+e_2 \neq 0$,} \\
\ProjMod{n_1+n_2,0}{\tn_1+\tn_2,0} & \text{if $e_1+e_2 = 0$,}
\end{cases}
\end{equation}
where $\ProjMod{n_1+n_2,0}{\tn_1+\tn_2,0}$ is an indecomposable $4$-dimensional module.  We remark that the average $N$-eigenvalue of the states of $\ProjMod{n_1+n_2,0}{\tn_1+\tn_2,0}$ is $n_1+n_2$, in accordance with its labels.  Its structure may be visualised as follows:
\begin{equation} \label{pic:AtypProj}
\parbox[c]{0.6\textwidth}{
\begin{center}
\begin{tikzpicture}[auto,thick]
\node (top) at (0,1.5) [] {$\ket{v} \otimes \psi^- \ket{w} - \psi^- \ket{v} \otimes \ket{w}$};
\node (left) at (-3.5,0) [] {$\ket{v} \otimes \ket{w}$};
\node (right) at (3.5,0) [] {$\psi^- \ket{v} \otimes \psi^- \ket{w}$};
\node (bot) at (0,-1.5) [] {$\ket{v} \otimes \psi^- \ket{w} + \psi^- \ket{v} \otimes \ket{w}$};
\draw [->] (top) to node {$\psi^+$} (left);
\draw [->] (top) to node [swap] {$\psi^-$} (right);
\draw [->] (left) to node {$\psi^-$} (bot);
\draw [->] (right) to node [swap] {$\psi^+$} (bot);
\end{tikzpicture}
\end{center}
}
.
\end{equation}
Note that $\ProjMod{n,0}{\tn,0}$ is not isomorphic to the Verma module $\VerMod{n,0}{\tn,0}$ as the $\tpsi^{\pm}$ act as zero and $\tN$ acts semisimply.  They do share the same composition factors:  $\AtypMod{n+1,0}{\tn,0}$, $\AtypMod{n,0}{\tn,0}$ (twice) and $\AtypMod{n-1,0}{\tn,0}$.  It also follows that the Casimirs $Q_1$ and $Q_2$ both act semisimply on $\ProjMod{n,0}{\tn,0}$.

The tensor product of a semitypical and a typical is easy to decompose:
\begin{equation} \label{TP:SxT}
\StypMod{n_1,e_1}{\tn_1,0} \otimes \TypMod{n_2,e_2}{\tn_2,\te_2} = \TypMod{n_1+n_2+1/2,e_1+e_2}{\tn_1+\tn_2,\te_2} \oplus \TypMod{n_1+n_2-1/2,e_1+e_2}{\tn_1+\tn_2,\te_2}.
\end{equation}
This follows because the tensor product of the \hwss{} generates a Verma module with typical $\tE$-eigenvalue $\te_2 \neq 0$.  Checking eigenvalues shows that it is therefore isomorphic to $\TypMod{n_1+n_2+1/2,e_1+e_2}{\tn_1+\tn_2,\te_2}$.  Similarly, the tensor product of the \lwss{} generates the other summand as a typical lowest weight Verma module.

It remains then to decompose the tensor product of two typicals $\TypMod{n_1,e_1}{\tn_1,\te_1}$ and $\TypMod{n_2,e_2}{\tn_2,\te_2}$.  When $\te_1 + \te_2 \neq 0$, the argument justifying \eqref{TP:SxT} allows us to deduce that the decomposition will involve the two typical irreducibles $\TypMod{n_1+n_2 \pm 1,e_1+e_2}{\tn_1+\tn_2,\te_1+\te_2}$.  The remainder of the decomposition is more interesting.  Searching for \hwss{} of $N$-eigenvalue $n_1+n_2+1$, we find only one:
\begin{equation}
\ket{\chi} = \te_1 \ket{v} \otimes \tpsi^- \ket{w} - \te_2 \tpsi^- \ket{v} \otimes \ket{w}.
\end{equation}
Here, $\ket{v}$ and $\ket{w}$ denote the \hwss{} of $\TypMod{n_1,e_1}{\tn_1,\te_1}$ and $\TypMod{n_2,e_2}{\tn_2,\te_2}$, respectively.  There is, however, a generalised \hws{} in the sense that it only fails to be an eigenstate of $\tN$.  Indeed, we may choose it to be
\begin{equation}
\ket{\lambda} =
\te_2 \psi^- \ket{v} \otimes \ket{w} - \te_1 \ket{v} \otimes \psi^- \ket{w} +
e_2 \tpsi^- \ket{v} \otimes \ket{w} - e_1 \ket{v} \otimes \tpsi^- \ket{w},
\end{equation}
in which case we have $\psi^+ \ket{\lambda} = \tpsi^+ \ket{\lambda} = 0$ and $\tN \ket{\lambda} = \brac{\tn_1+\tn_2} \ket{\lambda} + \ket{\chi}$.  It follows that the (typical) decomposition for the tensor product of two typicals is
\begin{equation} \label{TP:TxT}
\TypMod{n_1,e_1}{\tn_1,\te_1} \otimes \TypMod{n_2,e_2}{\tn_2,\te_2} =
\TypMod{n_1+n_2+1,e_1+e_2}{\tn_1+\tn_2,\te_1+\te_2} \oplus
\GenTypMod{n_1+n_2,e_1+e_2}{\tn_1+\tn_2,\te_1+\te_2}{2} \oplus
\TypMod{n_1+n_2-1,e_1+e_2}{\tn_1+\tn_2,\te_1+\te_2} \qquad \text{($\te_1+\te_2 \neq 0$),}
\end{equation}
where $\GenTypMod{n,e}{\tn,\te}{2}$ denotes an $8$-dimensional indecomposable with the structure
\begin{equation} \label{picGenTyp2}
\parbox[c]{0.25\textwidth}{
\begin{center}
\begin{tikzpicture}[auto,thick]
\node (left) at (-1.5,0) [] {$\TypMod{n,e}{\tn,\te}$};
\node (right) at (1.5,0) [] {$\TypMod{n,e}{\tn,\te}$};
\draw [->,dotted] (left) to (right);
\end{tikzpicture}
\end{center}
}
,
\end{equation}
the dotted arrow again indicating a non-semisimple action of $\tN$.

In more detail, $\GenTypMod{n,e}{\tn,\te}{2}$ is freely generated by the action of $\psi^-$ and $\tpsi^-$ from two states $\ket{v}$ and $\ket{w}$ (say) which are related by $\tN \ket{v} = \tn \ket{v} + \ket{w}$.  It is therefore one of the generalised Verma modules that we discussed briefly in \secref{sec:FinGL11Reps}.  This implies that $\tN - \tn$ sends $\tpsi^- \ket{v}$ to $\tpsi^- \ket{w}$ and $\tpsi^- \psi^- \ket{v}$ to $\tpsi^- \psi^- \ket{w}$.  However,
\begin{equation}
\tbrac{\tN - \tn} \psi^- \ket{v} = \psi^- \ket{w} - \tpsi^- \ket{v}, \qquad \text{hence} \qquad \tbrac{\tN - \tn}^2 \psi^- \ket{v} = -2 \tpsi^- \ket{w}.
\end{equation}
Pictorially, we have the following:
\begin{equation} \label{picGenTyp2'}
\parbox[c]{0.6\textwidth}{
\begin{center}
\begin{tikzpicture}[auto,thick]
\node (top1) at (0,1.5) [] {$\ket{v}$};
\node (left1) at (-1.5,0) [] {$\psi^- \ket{v}$};
\node (right1) at (1.5,0) [] {$\Bigl\{ \tpsi^- \ket{v} \Bigr.$,};
\node (bot1) at (0,-1.5) [] {$\tpsi^- \psi^- \ket{v}$};
\node (top2) at (4.5,1.5) [] {$\ket{w}$};
\node (left2) at (3,0) [] {$\Bigl. \psi^- \ket{w} \Bigr\}$};
\node (right2) at (6,0) [] {$\tpsi^- \ket{w}$};
\node (bot2) at (4.5,-1.5) [] {$\tpsi^- \psi^- \ket{w}$};
\node (mid) at (2.25,0) [] {};
\draw [-] (top1) to (left1);
\draw [-] (top1) to (right1);
\draw [-] (left1) to (bot1);
\draw [-] (right1) to (bot1);
\draw [-] (top2) to (left2);
\draw [-] (top2) to (right2);
\draw [-] (left2) to (bot2);
\draw [-] (right2) to (bot2);
\draw [->,dotted] (left1) to (right1);
\draw [->,dotted] (left2) to (right2);
\draw [->,dotted] (top1) to (top2);
\draw [->,dotted] (bot1) to (bot2);
\end{tikzpicture}
\end{center}
}
.
\end{equation}
In particular, $\tN$ has a rank $3$ Jordan cell when acting on $\GenTypMod{n,e}{\tn,\te}{2}$.  We remark that the Casimirs \eqref{eqn:Casimirs} both act non-semisimply on $\GenTypMod{n,e}{\tn,\te}{2}$:
\begin{equation} \label{eqn:CasimirAction}
Q_1 \ket{v} = \tbrac{n \te + \tn e} \ket{v} + e \ket{w}, \qquad Q_2 \ket{v} = \tn \te \ket{v} + \te \ket{w}.
\end{equation}
Their Jordan cells are, however, limited to rank $2$.

It remains to analyse the case of typical tensor typical when $\te_1 + \te_2 = 0$, which then splits into two subcases according as to whether $e_1+e_2 = 0$ or not.  These cases are tedious and messy, so we omit a detailed description of the resulting decomposition into indecomposables.  It is, however, easy to determine the composition factors of the tensor product (the irreducible modules which are ``glued together'' in the result).  For $\TypMod{n_1,e_1}{\tn_1,\te_1} \otimes \TypMod{n_2,e_2}{\tn_2,\te_2}$ with $\te_1 + \te_2 = 0$, the composition factors are
\begin{equation}
\begin{cases}
\text{$\StypMod{n_1+n_2+3/2,e_1+e_2}{\tn_1+\tn_2,0}$, $3 \: \StypMod{n_1+n_2+1/2,e_1+e_2}{\tn_1+\tn_2,0}$, $3 \: \StypMod{n_1+n_2-1/2,e_1+e_2}{\tn_1+\tn_2,0}$, $\StypMod{n_1+n_2-3/2,e_1+e_2}{\tn_1+\tn_2,0}$} & \text{if $e_1+e_2 \neq 0$,} \\
\text{$\AtypMod{n_1+n_2+2,0}{\tn_1+\tn_2,0}$, $4 \: \AtypMod{n_1+n_2+1,0}{\tn_1+\tn_2,0}$, $6 \: \AtypMod{n_1+n_2,0}{\tn_1+\tn_2,0}$, $4 \: \AtypMod{n_1+n_2-1,0}{\tn_1+\tn_2,0}$, $\AtypMod{n_1+n_2-2,0}{\tn_1+\tn_2,0}$} & \text{if $e_1+e_2 = 0$,}
\end{cases}
\end{equation}
where the multiplicities of the factors, if greater than $1$, are indicated via coefficients.  Unfortunately, one still has the $\ProjMod{}{}$- and $\GenTypMod{}{}{2}$-type modules.  We have not attempted to understand their tensor product rules, though it is again easy to work out the composition factors of the results.  It seems likely that these decompositions will generate further indecomposables including generalised Verma modules ($\GenTypMod{}{}{m}$-type modules) in which $m$ identical Verma modules, for arbitrarily large $m$, are glued together by a non-semisimple action of $\tN$.  We therefore content ourselves with describing the general tensor product rules in the associated Grothendieck ring.  This is the quotient of the true representation ring in which indecomposables are identified with the direct sum of their composition factors.  Elements of this quotient will be indicated by square brackets.

\begin{prop} \label{prop:TPRules}
The product induced by the tensor product on the associated Grothendieck ring of $\DSLSA{gl}{1}{1}$ is given, on the basis of (equivalence classes of) irreducibles, by
\begin{equation}
\begin{gathered}
\sqbrac{\AtypMod{n_1,0}{\tn_1,0}} \otimes \sqbrac{\AtypMod{n_2,0}{\tn_2,0}} = \sqbrac{\AtypMod{n_1+n_2,0}{\tn_1+\tn_2,0}},
\qquad
\sqbrac{\AtypMod{n_1,0}{\tn_1,0}} \otimes \sqbrac{\StypMod{n_2,e_2}{\tn_2,0}} = \sqbrac{\StypMod{n_1+n_2,e_2}{\tn_1+\tn_2,0}},
\\
\sqbrac{\AtypMod{n_1,0}{\tn_1,0}} \otimes \sqbrac{\TypMod{n_2,e_2}{\tn_2,\te_2}} = \sqbrac{\TypMod{n_1+n_2,e_2}{\tn_1+\tn_2,\te_2}},
\qquad
\sqbrac{\StypMod{n_1,e_1}{\tn_1,0}} \otimes \sqbrac{\StypMod{n_2,e_2}{\tn_2,0}} = \sqbrac{\TypMod{n_1+n_2,e_1+e_2}{\tn_1+\tn_2,0}}
\\
\sqbrac{\StypMod{n_1,e_1}{\tn_1,0}} \otimes \sqbrac{\TypMod{n_2,e_2}{\tn_2,\te_2}} = \sqbrac{\TypMod{n_1+n_2+1/2,e_1+e_2}{\tn_1+\tn_2,\te_2}} \oplus \sqbrac{\TypMod{n_1+n_2-1/2,e_1+e_2}{\tn_1+\tn_2,\te_2}},
\\
\sqbrac{\TypMod{n_1,e_1}{\tn_1,\te_1}} \otimes \sqbrac{\TypMod{n_2,e_2}{\tn_2,\te}} = \sqbrac{\TypMod{n_1+n_2+1,e_1+e_2}{\tn_1+\tn_2,\te_1+\te_2}} \oplus 2 \: \sqbrac{\TypMod{n_1+n_2,e_1+e_2}{\tn_1+\tn_2,\te_1+\te_2}} \oplus \sqbrac{\TypMod{n_1+n_2-1,e_1+e_2}{\tn_1+\tn_2,\te_1+\te_2}}.
\end{gathered}
\end{equation}
\end{prop}

\noindent We remark that we have not fully decomposed these Grothendieck tensor product rules into equivalence classes of irreducibles.  Instead, we have expressed them in terms of those of the irreducible typicals $\tsqbrac{\TypMod{n,e}{\tn,\te}}$ and the \emph{reducible} Verma modules $\tsqbrac{\VerMod{n,e}{\tn,0}} = \tsqbrac{\TypMod{n,e}{\tn,0}}$, when convenient.  For example, $\tsqbrac{\StypMod{n_1,e_1}{\tn_1,0}} \otimes \tsqbrac{\StypMod{n_2,e_2}{\tn_2,0}}$ decomposes into four atypicals or two semitypicals according as to whether $e_1 + e_2$ vanishes or not.  It is tedious, but easy, to perform such full decompositions when required.

\section{The Takiff Superalgebra of $\AKMSA{gl}{1}{1}$} \label{sec:ExtGL11}

\subsection{Algebraic Preliminaries}

The Lie superalgebra $\SLSA{gl}{1}{1}$ is not simple (it has a non-trivial centre spanned by $E$), but it possesses an affinisation $\AKMSA{gl}{1}{1}$ because there is a non-degenerate even supersymmetric bilinear form $\killing{\cdot}{\cdot}$ given by the supertrace of the product in the defining $\left( 1 \middle\vert 1 \right)$-dimensional representation:
\begin{equation}
N = \frac{1}{2}
\begin{pmatrix}
1 & 0 \\
0 & -1
\end{pmatrix}
, \qquad E =
\begin{pmatrix}
1 & 0 \\
0 & 1
\end{pmatrix}
, \qquad \psi^+ =
\begin{pmatrix}
0 & 1 \\
0 & 0
\end{pmatrix}
, \qquad \psi^- =
\begin{pmatrix}
0 & 0 \\
1 & 0
\end{pmatrix}
.
\end{equation}
The affinisation is then as in \eqnref{eqn:DefAff}.  The Takiff superalgebra of $\AKMSA{gl}{1}{1}$ thus has the following non-vanishing commutation relations:
\begin{equation}
\begin{aligned}
\comm{N_r}{E_s} &= r k \delta_{r+s,0}, \\
\comm{\tN_r}{E_s} &= r \tk \delta_{r+s,0}, \\
\comm{N_r}{\tE_s} &= r \tk \delta_{r+s,0},
\end{aligned}
\qquad
\begin{aligned}
\comm{N_r}{\psi^{\pm}_s} = \pm \psi^{\pm}_{r+s}, \\
\comm{\tN_r}{\psi^{\pm}_s} = \pm \tpsi^{\pm}_{r+s}, \\
\comm{N_r}{\tpsi^{\pm}_s} = \pm \tpsi^{\pm}_{r+s},
\end{aligned}
\qquad
\begin{aligned}
\acomm{\psi^+_r}{\psi^-_s} = E_{r+s} + r k \delta_{r+s,0}, \\
\acomm{\tpsi^+_r}{\psi^-_s} = \tE_{r+s} + r \tk \delta_{r+s,0}, \\
\acomm{\psi^+_r}{\tpsi^-_s} = \tE_{r+s} + r \tk \delta_{r+s,0}.
\end{aligned}
\end{equation}
We denote this Takiff superalgebra by $\DAKMSA{gl}{1}{1}$.

As usual, the triangular decomposition \eqref{eqn:FinGL11TriDec} lifts to one for $\DAKMSA{gl}{1}{1}$, so we may define \hwss{} and Verma modules.  Explicitly, we take a \hws{} to be an eigenvector of $N_0$, $E_0$, $\tN_0$ and $\tE_0$ which is annihilated by $\psi^+_0$, $\tpsi^+_0$ and every mode with a positive index.  Because $\SLSA{gl}{1}{1}$ is not simple, \thmref{thm:ExtSugawara} does not apply.  However, it is not hard to check that there is a unique energy momentum tensor coming from the Casimirs \eqref{eqn:Casimirs}:
\begin{multline} \label{eqn:DefTGL11}
\func{T}{z} = \frac{1}{\tk} \sqbrac{\normord{\func{N}{z} \func{\tE}{z}} + \normord{\func{E}{z} \func{\tN}{z}} - \normord{\func{\psi^+}{z} \func{\tpsi^-}{z}} + \normord{\func{\psi^-}{z} \func{\tpsi^+}{z}}} \\
- \frac{k}{\tk^2} \sqbrac{\normord{\func{\tN}{z} \func{\tE}{z}} - \normord{\func{\tpsi^+}{z} \func{\tpsi^-}{z}}} + \frac{1}{\tk^2} \normord{\func{\tE}{z} \func{\tE}{z}}.
\end{multline}
The currents of $\DAKMSA{gl}{1}{1}$ are dimension $1$ primary fields with respect to this tensor and the central charge is $c=0$, which agrees with \eqnref{eqn:DoubledC}.  It follows that the conformal dimension of a \hws{} whose weight is $\tbrac{n,e,\tn,\te}$ (these are the eigenvalues of $N_0$, $E_0$, $\tN_0$ and $\tE_0$, respectively) is given by
\begin{equation}
\Delta = \brac{n-1} \frac{\te}{\tk} + \frac{k}{\tk} \: \tn \brac{\frac{e}{k} - \frac{\te}{\tk}} + \frac{\te^2}{\tk^2}.
\end{equation}
If we let $\AffVerMod{n,e}{\tn,\te}$ denote the Verma module generated by a \hws{} of weight $\tbrac{n+1,e,\tn,\te}$, so that $n$ labels the average $N_0$-eigenvalue of the states of minimal conformal dimension, then this minimal conformal dimension will be
\begin{equation} \label{eqn:GL11ConfDim}
\Delta^{n,e}_{\tn,\te} = n \: \frac{\te}{\tk} + \frac{k}{\tk} \: \tn \brac{\frac{e}{k} - \frac{\te}{\tk}} + \frac{\te^2}{\tk^2}.
\end{equation}

We remark that when $k$ and $\tk$ are non-vanishing, we are free to rescale the generators as follows:
\begin{equation} \label{eqn:Rescaling}
\begin{aligned}
N_r &\longrightarrow N_r, \vphantom{\frac{1}{\sqrt{k}}} \\
\tN_r &\longrightarrow \frac{k}{\tk} \tN_r, \vphantom{\frac{\sqrt{k}}{\tk}}
\end{aligned}
\qquad
\begin{aligned}
E_r &\longrightarrow \frac{1}{k} E_r, \vphantom{\frac{1}{\sqrt{k}}} \\
\tE_r &\longrightarrow \frac{1}{\tk} \tE_r, \vphantom{\frac{\sqrt{k}}{\tk}}
\end{aligned}
\qquad
\begin{aligned}
\psi^{\pm}_r &\longrightarrow \frac{1}{\sqrt{k}} \psi^{\pm}_r, \\
\tpsi^{\pm}_r &\longrightarrow \frac{\sqrt{k}}{\tk} \tpsi^{\pm}.
\end{aligned}
\end{equation}
The commutation relations of the rescaled generators now have $k = \tk = 1$.  We will not be interested in the critical case $\tk = 0$ when $\func{T}{z}$ fails to exist and will generally ignore the case $k=0$, $\tk \neq 0$ which may well be of interest.  We note however that when $k=0$ and $\tk \neq 0$, the rescaling
\begin{equation}
\tN_r \longrightarrow \frac{1}{\tk} \tN_r, \qquad \tE_r \longrightarrow \frac{1}{\tk} \tE_r, \qquad \tpsi_r \longrightarrow \frac{1}{\tk} \tpsi_r
\end{equation}
will set $\tk$ to $1$.  While it would be very convenient, we shall not assume in what follows that $k$ and $\tk$ have been rescaled as above.  However, a consequence of this potential for rescaling is that the natural units for measuring the eigenvalues of $E$, $\tN$ and $\tE$ are $k$, $\tk / k$ and $\tk$, respectively.  This is nicely illustrated in \eqnref{eqn:GL11ConfDim} and we find it convenient to use these units as indicators that the algebra which follows is correct.

\subsection{Representations and Characters} \label{sec:AffGL11Reps}

One immediate consequence of the formula \eqref{eqn:GL11ConfDim} is that when $\te = 0$, every singular vector of $\AffVerMod{n,e}{\tn,0}$ occurs at the zeroth grade.  This follows from the fact that singular vectors are highest weight, so their conformal dimensions are given by
\begin{equation} \label{eqn:ConfDimTE=0}
\Delta^{n',e}_{\tn,0} = \frac{\tn e}{\tk},
\end{equation}
and the fact that the weight of such a singular vector can only differ from that of the (generating) \hws{} in its $N_0$-eigenvalue (moreover, $n' - n \in \ZZ$).  In fact, we may now conclude from \secref{sec:FinGL11Reps} that the maximal submodule of $\AffVerMod{n,e}{\tn,0}$ will be generated by the unique non-trivial singular vector (up to scalar multiples) whose grade is $0$ and $N_0$-eigenvalue is $n$, provided that $e \neq 0$.  In particular, the irreducible quotient will have two independent zero-grade states, one of $N_0$-eigenvalue $n+1$ and one of $N_0$-eigenvalue $n$.  We shall therefore refer to it as semitypical and denote it by $\AffStypMod{n+1/2,e}{\tn,0}$.  In the case $e=0$, the maximal submodule will be generated by a generalised singular vector of grade $0$ and $N_0$-eigenvalue $n$ and the irreducible quotient, denoted by $\AffAtypMod{n+1,0}{\tn,0}$, will have a one-dimensional zero-grade subspace where the $N_0$-eigenvalue is $n+1$.  This irreducible will be said to be atypical.

Consider therefore a Verma module $\AffVerMod{n,e}{\tn,\te}$ with $\te \neq 0$.  From the result of \propref{prop:FinGL11Irreps}, there are no non-trivial singular vectors among the zero-grade states.  Suppose then that there exists another singular vector in $\AffVerMod{n,e}{\tn,\te}$ with weight $\tbrac{n + \nu,e,\tn,\te}$ and conformal dimension $\Delta^{n + \nu,e}_{\tn,\te} = \Delta^{n,e}_{\tn,\te} + m$, where $\nu \in \ZZ$ and $m$ is a positive integer.  It follows that
\begin{equation}
m = \Delta^{n + \nu,e}_{\tn,\te} - \Delta^{n,e}_{\tn,\te} = \nu \: \frac{\te}{\tk}.
\end{equation}
However, a quick look at the Poincar\'{e}-Birkhoff-Witt basis indicates that the quantities $\nu$ and $m$ are also restricted to satisfy $\abs{\nu} \leqslant m+1$, hence
\begin{equation} \label{eqn:Constraint}
m = \abs{m} = \abs{\nu} \abs{\frac{\te}{\tk}} \leqslant \brac{m+1} \abs{\frac{\te}{\tk}}.
\end{equation}
If we restrict $\te$ so that $0 < \tabs{\te / \tk} \leqslant \tfrac{1}{2}$, then the only positive integer $m$ satisfying \eqref{eqn:Constraint} will be $m=1$, which requires that $\te / \tk = \pm \tfrac{1}{2}$ and $\nu = \pm 2$.  But, such a singular vector must have the form $\tbrac{\alpha \psi^{\pm}_{-1} + \beta \tpsi^{\pm}_{-1}} \ket{v}$, where $\ket{v}$ is either the \hws{} (if the fermions both have ``$+$'' labels) or its $\psi^-_0 \tpsi^-_0$-descendant (if the labels are both ``$-$''), and it is easy to show that such a vector is not singular for $\te / \tk = \pm \tfrac{1}{2}$.  This then proves that the Verma modules $\AffVerMod{n,e}{\tn,\te}$ with $0 < \tabs{\te / \tk} \leqslant \tfrac{1}{2}$ have no non-trivial singular vectors and so are irreducible.  We therefore regard them as typical and denote them by $\AffTypMod{n,e}{\tn,\te}$.  To summarise:

\begin{lem} \label{lem:VermaFundDom}
The Verma modules $\AffVerMod{n,e}{\tn,\te}$ of $\DAKMSA{gl}{1}{1}$ are irreducible when $0 < \tabs{\te / \tk} \leqslant \tfrac{1}{2}$.  Contrarily, the Verma module $\AffVerMod{n,e}{\tn,0}$ is reducible and its maximal submodule is generated by the $\tpsi^-_0$-descendant of the (generating) \hws{} if $e \neq 0$, and by the $\psi^-_0$-descendant of the (generating) \hws{} if $e=0$.
\end{lem}

The characters of the irreducible $\DAKMSA{gl}{1}{1}$-modules with $\tabs{\te / \tk} \leqslant \tfrac{1}{2}$ are now easy to write down.  We will, for now, define characters so as to keep track of the $N_0$-eigenvalues and conformal dimensions of the states of the representations, the eigenvalues of $E_0$, $\tN_0$ and $\tE_0$ being constant for any given indecomposable:
\begin{equation}
\fch{\mathcal{M}}{z;q} = \traceover{\mathcal{M}} z^{N_0} q^{L_0 - c/24}.
\end{equation}
This definition of character will be generalised when we consider modular properties in \secref{sec:GL11Mod}.  Remembering that $c=0$, the characters of the typical, semitypical and atypical irreducibles with $\tabs{\te / \tk} \leqslant \tfrac{1}{2}$ are therefore given by
\begin{equation} \label{ch:GL11}
\begin{aligned}
\ch{\AffTypMod{n,e}{\tn,\te}} &= z^{n+1} q^{\Delta^{n,e}_{\tn,\te}} \prod_{i=1}^{\infty} \frac{\brac{1+zq^i}^2 \brac{1+z^{-1}q^{i-1}}^2}{\brac{1-q^i}^4} & &\text{($0 < \tabs{\te / \tk} \leqslant \tfrac{1}{2}$),} \\
\ch{\AffStypMod{n,e}{\tn,0}} &= z^{n+1/2} q^{\tn e / \tk} \prod_{i=1}^{\infty} \frac{\brac{1+zq^i}^2 \brac{1+z^{-1}q^i} \brac{1+z^{-1}q^{i-1}}}{\brac{1-q^i}^4} & &\text{($e \neq 0$),} \\
\ch{\AffAtypMod{n,0}{\tn,0}} &= z^n \prod_{i=1}^{\infty} \frac{\brac{1+zq^i}^2 \brac{1+z^{-1}q^i}^2}{\brac{1-q^i}^4}. & &
\end{aligned}
\end{equation}
Here, we remark that \eqref{eqn:ConfDimTE=0} gives $\Delta^{n,e}_{\tn,0} = \tn e / \tk$ and $\Delta^{n,0}_{\tn,0} = 0$.

To extend these character formulae beyond $\tabs{\te / \tk} \leqslant \tfrac{1}{2}$, we employ spectral flow automorphisms. Their effectiveness in classifying irreducible modules is well known for affine algebras and superalgebras \cite{FeiEqu98,SalGL106}. These automorphisms do not preserve the grading operator $L_0$ and usually arise through an appropriate shifting of the mode indices.  The spectral flow automorphisms that are useful here are those generated by $\tsigma$, where
\begin{equation}
\begin{aligned}
\tfunc{\tsigma}{\psi^{\pm}_r} &= \psi^{\pm}_{r \mp 1}, \\
\tfunc{\tsigma}{\tpsi^{\pm}_r} &= \tpsi^{\pm}_{r \mp 1},
\end{aligned}
\qquad
\begin{aligned}
\tfunc{\tsigma}{N_r} &= N_r, \\
\tfunc{\tsigma}{\tN_r} &= \tN_r,
\end{aligned}
\qquad
\begin{aligned}
\tfunc{\tsigma}{E_r} &= E_r - \delta_{r,0} k, \\
\tfunc{\tsigma}{\tE_r} &= \tE_r - \delta_{r,0} \tk.
\end{aligned}
\end{equation}
It is easy to check that $\tsigma$ preserves $k$ and $\tk$, but a little tedious to check from \eqnref{eqn:DefTGL11} that
\begin{equation}
\tfunc{\tsigma}{L_0} = L_0 - N_0.
\end{equation}
Automorphisms may be used to twist the action of an algebra, giving rise to new modules.  In particular, if $\ket{v}$ is a state of a $\DAKMSA{gl}{1}{1}$-module $\mathcal{M}$, then we may define $\tsigma \ket{v}$ to be a state of a module $\sfmod{}{\mathcal{M}}$, the \emph{spectral flow image} of $\mathcal{M}$, upon which $\DAKMSA{gl}{1}{1}$ acts through
\begin{equation}
J \cdot \tsigma \ket{v} = \tsigma \brac{\tfunc{\tsigma^{-1}}{J} \ket{v}} \qquad \text{($J \in \DAKMSA{gl}{1}{1}$).}
\end{equation}
It is a good exercise to check that taking $\ket{v}$ to have weight $\tbrac{n,e,\tn,\te}$ and conformal dimension $\Delta$ leads to $\tsigma \ket{v}$ having weight $\tbrac{n,e+k,\tn,\te+\tk}$ and conformal dimension $\Delta + n$.  This leads quickly to the character relations
\begin{equation} \label{eqn:GL11SFChar}
\fch{\sfmod{}{\mathcal{M}}}{z;q} = \fch{\mathcal{M}}{zq;q}.
\end{equation}

Being an automorphism, spectral flow preserves module structure.  In particular, it maps irreducibles to irreducibles.  Consider the spectral flow of a typical module $\AffTypMod{n,e}{\tn,\te}$ with $0 < \tabs{\te / \tk} \leqslant \tfrac{1}{2}$.  By \eqref{ch:GL11} and \eqref{eqn:GL11SFChar}, the character of the result is given by
\begin{align}
\fch{\sfmod{}{\AffTypMod{n,e}{\tn,\te}}}{z;q} &= z^{n+1} q^{n+1} q^{\Delta^{n,e}_{\tn,\te}} \prod_{i=1}^{\infty} \frac{\brac{1+zq^{i+1}}^2 \brac{1+z^{-1}q^{i-2}}^2}{\brac{1-q^i}^4} \notag \\
&= z^{n+1} q^{\Delta^{n,e}_{\tn,\te} + n + 1} \brac{\frac{1+z^{-1}q^{-1}}{1+zq}}^2 \prod_{i=1}^{\infty} \frac{\brac{1+zq^i}^2 \brac{1+z^{-1}q^{i-1}}^2}{\brac{1-q^i}^4} \notag \\
&= z^{n-1} q^{\Delta^{n-2,e+k}_{\tn,\te+\tk}} \prod_{i=1}^{\infty} \frac{\brac{1+zq^i}^2 \brac{1+z^{-1}q^{i-1}}^2}{\brac{1-q^i}^4} = \fch{\AffVerMod{n-2,e+k}{\tn,\te+\tk}}{z;q}.
\end{align}
The module $\sfmod{}{\AffTypMod{n,e}{\tn,\te}}$ is therefore an irreducible with the same character as $\AffVerMod{n-2,e+k}{\tn,\te+\tk}$.  To verify this identification, we only need to ascertain that the spectral flow module possesses a \hws{}.  But, if $\ket{v}$ is the \hws{} of $\AffTypMod{n,e}{\tn,\te}$, it is easy to check that
\begin{equation}
\tpsi^-_1 \psi^-_1 \tsigma \ket{v} = \tsigma \brac{\tpsi^-_0 \psi^-_0 \ket{v}}
\end{equation}
is a \hws{} of $\sfmod{}{\AffTypMod{n,e}{\tn,\te}}$.  It follows, by irreducibility, that this is \emph{the} \hws{} of $\sfmod{}{\AffTypMod{n,e}{\tn,\te}}$, that $\sfmod{}{\AffTypMod{n,e}{\tn,\te}}$ is isomorphic to $\AffVerMod{n-2,e+k}{\tn,\te+\tk}$, and that $\AffVerMod{n-2,e+k}{\tn,\te+\tk}$ is an irreducible Verma module.  Iterating these conclusions appropriately, we arrive at the following result:

\begin{prop} \label{prop:TypChar}
The Verma modules $\AffVerMod{n,e}{\tn,\te}$ of $\DAKMSA{gl}{1}{1}$ with $\te / \tk \notin \ZZ$ are irreducible.  We therefore refer to them as \emph{typical} and denote them by $\AffTypMod{n,e}{\tn,\te}$.  Their characters are given by
\begin{equation} \label{ch:GL11Typ}
\ch{\AffTypMod{n,e}{\tn,\te}} = z^{n+1} q^{\Delta^{n,e}_{\tn,\te}} \prod_{i=1}^{\infty} \frac{\brac{1+zq^i}^2 \brac{1+z^{-1}q^{i-1}}^2}{\brac{1-q^i}^4} \qquad \text{($\te / \tk \notin \ZZ$)}
\end{equation}
and they are related under spectral flow by
\begin{equation} \label{eqn:SFTyp}
\sfmod{\ell}{\AffTypMod{n,e}{\tn,\te}} \cong \AffTypMod{n - 2 \ell , e + \ell k}{\tn , \te + \ell \tk}.
\end{equation}
\end{prop}

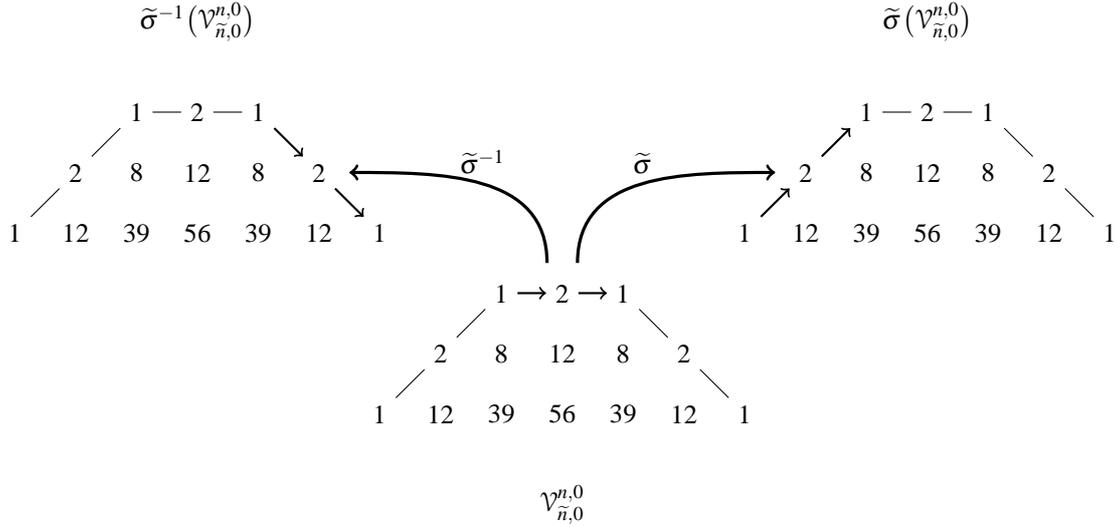
\begin{figure}
\begin{center}
\begin{tikzpicture}[scale=0.8]
\node at (0,-1.5) {$\AffVerMod{n,0}{\tn,0}$};
\node (a) at (-3,0) {1};
\node at (-2,0) {12};
\node at (-1,0) {39};
\node at (0,0) {56};
\node at (1,0) {39};
\node at (2,0) {12};
\node (g) at (3,0) {1};
\node (b) at (-2,1) {2};
\node at (-1,1) {8};
\node at (0,1) {12};
\node at (1,1) {8};
\node (f) at (2,1) {2};
\node (c) at (-1,2) {1};
\node (d) at (0,2) {2};
\node (e) at (1,2) {1};
\draw (a) -- (b) -- (c);
\draw (e) -- (f) -- (g);
\draw [->,thick] (c) -- (d);
\draw [->,thick] (d) -- (e);
\begin{scope}[shift={(6,3)}]
\node at (0,3.5) {$\sfmod{}{\AffVerMod{n,0}{\tn,0}}$};
\node (a) at (-3,0) {1};
\node at (-2,0) {12};
\node at (-1,0) {39};
\node at (0,0) {56};
\node at (1,0) {39};
\node at (2,0) {12};
\node (g) at (3,0) {1};
\node (b) at (-2,1) {2};
\node at (-1,1) {8};
\node at (0,1) {12};
\node at (1,1) {8};
\node (f) at (2,1) {2};
\node (c) at (-1,2) {1};
\node (d) at (0,2) {2};
\node (e) at (1,2) {1};
\draw (c) -- (d) -- (e) -- (f) -- (g);
\draw [->,thick] (a) -- (b);
\draw [->,thick] (b) -- (c);
\end{scope}
\begin{scope}[shift={(-6,3)}]
\node at (0,3.5) {$\sfmod{-1}{\AffVerMod{n,0}{\tn,0}}$};
\node (a) at (-3,0) {1};
\node at (-2,0) {12};
\node at (-1,0) {39};
\node at (0,0) {56};
\node at (1,0) {39};
\node at (2,0) {12};
\node (g) at (3,0) {1};
\node (b) at (-2,1) {2};
\node at (-1,1) {8};
\node at (0,1) {12};
\node at (1,1) {8};
\node (f) at (2,1) {2};
\node (c) at (-1,2) {1};
\node (d) at (0,2) {2};
\node (e) at (1,2) {1};
\draw (a) -- (b) -- (c) -- (d) -- (e);
\draw [->,thick] (e) -- (f);
\draw [->,thick] (f) -- (g);
\end{scope}
\draw [->,very thick] (0.25,2.5) .. controls (0.25,4) and (2,4) .. (3.5,4) node [pos=0.5,above] {$\tsigma$};
\draw [->,very thick] (-0.25,2.5) .. controls (-0.25,4) and (-2,4) .. (-3.5,4) node [pos=0.5,above] {$\tsigma^{-1}$};
\end{tikzpicture}
\caption{How spectral flow acts on (atypical) Verma modules:  The numbers denote the multiplicities of the weight spaces of the Verma module (only the first few grades are shown) and the multiplicities are arranged so that the $N_0$-eigenvalue decreases from left to right and the $L_0$-eigenvalue increases from bottom to top.  The arrows, as in $a \to b$, indicate that the algebra action can map the states in the weight space labelled by $a$ to those labelled by $b$, \emph{but not vice-versa}.  For $\AffVerMod{n,0}{\tn,0}$, the top-left $1$ corresponds to the \hws{} whereas the top-right $1$ corresponds to the singular vector generating the irreducible submodule.} \label{fig:SF}
\end{center}
\end{figure}

To generalise this to the non-typical r\'{e}gimes which will presumably correspond to $\te / \tk \in \ZZ$, we take a closer look at the spectral flow images of the Verma modules with $\te = 0$.  A slight refinement of the argument used in the typical case leads us to the following conclusion:  When $\ell = -1, -2, -3, \ldots$, the spectral flow image $\sfmod{\ell}{\AffVerMod{n,e}{\tn,0}}$ has four (independent) zero-grade states whose projections are linearly independent in the irreducible quotient.  When $\ell = 1, 2, 3, \ldots$ however, $\sfmod{\ell}{\AffVerMod{n,e}{\tn,0}}$ likewise has four linearly independent zero-grade states, but they generate an irreducible submodule.  This is best understood by referring to \figref{fig:SF} which illustrates the action of spectral flow on the multiplicities of the weight spaces of an atypical Verma module (the illustration for semitypical Verma modules is practically identical).

For $e \neq 0$, the maximal submodule of $\AffVerMod{n,e}{\tn,0}$ is irreducible and it is isomorphic to the semitypical $\AffStypMod{n-1/2,e}{\tn,0}$.  Moreover, the quotient by this is isomorphic to the irreducible $\AffStypMod{n+1/2,e}{\tn,0}$.  As spectral flow respects module structure, we conclude that the four zero-grade states of $\sfmod{\ell}{\AffVerMod{n,e}{\tn,0}}$ generate $\sfmod{\ell}{\AffStypMod{n-1/2,e}{\tn,0}}$ for $\ell > 0$ and $\sfmod{\ell}{\AffStypMod{n+1/2,e}{\tn,0}}$ for $\ell < 0$.  As the average $N_0$-eigenvalue of the four zero-grade states is $n - 2 \ell$, we may extend the definition of semitypical irreducibles to $\te / \tk \in \ZZ$ via
\begin{equation} \label{eqn:SFStyp}
\AffStypMod{n,e}{\tn,\ell \tk} =
\begin{cases}
\sfmod{\ell}{\AffStypMod{n + 1/2 + 2 \ell , e - \ell k}{\tn,0}} & \text{if $e \neq \ell k$ and $\ell = -1, -2, -3, \ldots$} \\
\sfmod{\ell}{\AffStypMod{n - 1/2 + 2 \ell , e - \ell k}{\tn,0}} & \text{if $e \neq \ell k$ and $\ell = +1, +2, +3, \ldots$}
\end{cases}
\end{equation}
The condition for semitypical $\DAKMSA{gl}{1}{1}$-modules is thus $\te / \tk \in \ZZ$ and $e/k \neq \te / \tk$.

Finally, the definition of atypical irreducibles follows a similar pattern.  There is a unique irreducible submodule of $\AffVerMod{n,0}{\tn,0}$ and it is isomorphic to the atypical $\AffAtypMod{n-1,0}{\tn,0}$.  Furthermore, the quotient by the maximal submodule is isomorphic to $\AffAtypMod{n+1,0}{\tn,0}$.  We therefore define
\begin{equation} \label{eqn:SFAtyp}
\AffAtypMod{n,\ell k}{\tn,\ell \tk} =
\begin{cases}
\sfmod{\ell}{\AffAtypMod{n + 1 + 2 \ell , 0}{\tn,0}} & \text{if $\ell = -1, -2, -3, \ldots$} \\
\sfmod{\ell}{\AffAtypMod{n - 1 + 2 \ell , 0}{\tn,0}} & \text{if $\ell = +1, +2, +3, \ldots$}
\end{cases}
\end{equation}
Atypical $\DAKMSA{gl}{1}{1}$-modules thus require $e/k = \te / \tk \in \ZZ$.  We can now compute character formulae for the semitypical and atypical irreducibles in principle.  However, we shall not need these formulae in what follows.  Instead, we shall derive character formulae in which the characters are expressed as (infinite) linear combinations of typical characters.

\subsection{Modular Transformations} \label{sec:GL11Mod}

To investigate the modular transformation properties of the $\DAKMSA{gl}{1}{1}$ characters, we redefine them to depend on a much larger set of variables:
\beq
\fch{\mathcal{M}}{x,y,z;\tx,\ty,\tz;q} = \traceover{\mathcal{M}} x^k \tx^{\tk} y^{E_0} \ty^{\tE_0} z^{N_0} \tz^{\tN_0} q^{L_0 - c/24}.
\eeq
In fact, we will instead consider the supercharacters $\sch{\mathcal{M}}$ in which the trace is replaced by the supertrace.  With the parametrisation one should expect for these variables,
\beq \label{eqn:ModVars}
x = e^{2 \pi \ii t}, \quad \tx = e^{2 \pi \ii \ttt}, \quad y = e^{2 \pi \ii \mu}, \quad \ty = e^{2 \pi \ii \tmu}, \quad z = e^{2 \pi \ii \nu}, \quad \tz = e^{2 \pi \ii \tnu}, \quad q = e^{2 \pi \ii \tau},
\eeq
the $\SLG{SL}{2 ; \ZZ}$ generators $\modS$ and $\modT$ may be taken to act on the set of character variables as follows:
\begin{equation}
\begin{aligned}
\modS &\colon \modarg{t}{\mu}{\nu}{\ttt}{\tmu}{\tnu}{\tau} \longmapsto \modarg{t - \frac{\mu \nu}{\tau}}{\frac{\mu}{\tau}}{\frac{\nu}{\tau}}{\ttt - \frac{\tmu \nu}{\tau} - \frac{\mu \tnu}{\tau}}{\frac{\tmu}{\tau}}{\frac{\tnu}{\tau}}{-\frac{1}{\tau}}, \\
\modT &\colon \modarg{t}{\mu}{\nu}{\ttt}{\tmu}{\tnu}{\tau} \longmapsto \modarg{t}{\mu}{\nu}{\ttt}{\tmu}{\tnu}{\tau + 1}.
\end{aligned}
\end{equation}
One can easily check that these transformations satisfy $\modS^4 = \brac{\modS \modT}^6 = \id$.  A justification of the (perhaps unusual) transformations of $t$ and $\ttt$ will have to wait until the S-transformations of the supercharacters have been derived.

We begin with the typical supercharacters.  Assuming that the \hws{} is taken to be even (an odd \hws{} would just change the result by an overall sign), the supercharacter of $\AffTypMod{n,e}{\tn,\te}$ is obtained from \propref{prop:TypChar} by replacing $z$ by $-z$ in the infinite product and multiplying by the variables neglected there.  Thus,
\begin{align} \label{eqn:TypChar}
\sch{\AffTypMod{n,e}{\tn,\te}} &= x^k \tx^{\tk} y^e \ty^{\te} z^{n+1} \tz^{\tn} q^{\Delta^{n,e}_{\tn,\te}} \prod_{i=1}^{\infty} \frac{\brac{1-zq^i}^2 \brac{1-z^{-1}q^{i-1}}^2}{\brac{1-q^i}^4} \notag \\
&= \ii x^k y^e z^n \tx^{\tk} \ty^{\te} \tz^{\tn} q^{\brac{n \te + \tn e} / \tk + \brac{\te^2 - k \tn \te} / \tk^2} \frac{\Jth{1}{z;q}^2}{\func{\eta}{q}^6}.
\end{align}
The reason for considering supercharacters rather than characters is now clear.  The theta function $\jth{1}$ is transformed into itself under modular transformations, whereas the other theta functions transform into one another.  It follows that if we had taken characters rather than supercharacters, then we would have had to also consider \emph{twisted} characters and supercharacters which are related to ordinary characters and supercharacters by half-integer spectral flows.  We do not need this level of complexity for the application at hand (the Verlinde formula), but the methodology employed here works just as well for characters and twisted (super)characters.

Substituting \eqref{eqn:ModVars} into \eqref{eqn:TypChar}, we arrive at the following form for the typical supercharacters:
\begin{equation}
\sch{\AffTypMod{n,e}{\tn,\te}} = \ii \ee^{2 \pi \ii \brac{kt + \tk \ttt}} \ee^{2 \pi \ii \brac{e \mu + n \nu + \te \tmu + \tn \tnu}} \ee^{2 \pi \ii \sqbrac{\brac{n \te + \tn e} / \tk + \brac{\te^2 - k \tn \te} / \tk^2} \tau} \frac{\Jth{1}{\nu \vert \tau}^2}{\func{\eta}{\tau}^6}.
\end{equation}
The T-transformation is now immediate, giving
\begin{equation}
\modT \set{\sch{\AffTypMod{n,e}{\tn,\te}}} = \ee^{2 \pi \ii \Delta^{n,e}_{\tn,\te}} \sch{\AffTypMod{n,e}{\tn,\te}},
\end{equation}
as one expects.  The key to deriving the S-transformation is to try to write it as a Fourier transform (with an unknown bilinear form).  The result is as follows:

\begin{thm} \label{thm:TypMod}
The S-transformations of the typical $\DAKMSA{gl}{1}{1}$-supercharacters are given by
\begin{subequations}
\begin{equation}
\modS \set{\sch{\AffTypMod{n,e}{\tn,\te}}} = \underset{\RR^4}{\iiiint} \SMat{n,e}{\tn,\te}{n',e'}{\tn',\te'} \sch{\AffTypMod{n',e'}{\tn',\te'}} \: \dd n' \: \dd \brac{e'/k} \: \dd \brac{k \tn' / \tk} \: \dd \brac{\te' / \tk},
\end{equation}
where the S-matrix ``entries'' are given by
\begin{equation} \label{eqn:SMatTyp}
\SMat{n,e}{\tn,\te}{n',e'}{\tn',\te'} = \ee^{-2 \pi \ii \sqbrac{\brac{n \te' + n' \te + \tn e' + \tn' e} / \tk + \brac{2 \te \te' - k \tn \te' - k \tn' \te} / \tk^2}}.
\end{equation}
\end{subequations}
\end{thm}

\noindent The proof is a straight-forward, though rather tedious, verification.  One should evaluate the integrals in the order given (so the integral over $\te'$ should be performed first) using the analytic continuation of
\begin{equation}
\int_{\RR} e^{-ax^2 + bx} \: \dd x = \sqrt{\frac{\pi}{a}} \ee^{b^2 / 4a} \qquad \text{($\Re a > 0$)}
\end{equation}
to all $a \in \CC$.  In the course of this verification, the transformations of the variables $t$ and $\ttt$ are explained --- they must be chosen to transform in this manner so as to ``sop up'' unwanted phases arising from the transformations of the other variables.  This behaviour is, of course, identical to that of the analogue of $t$ and $\ttt$ used when considering ordinary affine Lie algebra characters.

We need to generalise this to the semitypical and atypical characters.  We will do this by deriving resolutions for the corresponding modules from which we will obtain the non-typical characters as infinite linear combinations of reducible Verma module characters.  These Verma module characters have the same form as those of the typical irreducibles, so \thmref{thm:TypMod} applies to them.  In this way, we construct the non-typical irreducible characters and deduce their S-matrix entries.  More precisely, this amounts to taking a (topological) basis of characters consisting of the typical characters and the reducible Verma module characters (which have the same form as the typical characters).  The S-matrix entries which we deduce for the non-typical modules will then describe the decomposition of an S-transformed non-typical character into this basis of typical and Verma module characters.

To begin, recall that when $e \neq 0$, the maximal submodule of the Verma module $\AffVerMod{n,e}{\tn,0}$ is isomorphic to the semitypical $\AffStypMod{n-1/2,e}{\tn,0}$ and that the corresponding quotient is isomorphic to the semitypical $\AffStypMod{n+1/2,e}{\tn,0}$.  This is conveniently summarised in the short exact sequence
\begin{equation}
\ses{\AffStypMod{n-1,e}{\tn,0}}{\AffVerMod{n-1/2,e}{\tn,0}}{\AffStypMod{n,e}{\tn,0}}
\end{equation}
(in which we have shifted indices slightly for convenience).  The sequence with $n$ replaced by $n-1$ will then have quotient (rightmost entry) matching the submodule (leftmost entry) of the original sequence.  It is easy to check that we can therefore \emph{splice} the two short exact sequences together and obtain a four-term exact sequence:
\begin{equation}
\ftes{\AffStypMod{n-2,e}{\tn,0}}{\AffVerMod{n-3/2,e}{\tn,0}}{\AffVerMod{n-1/2,e}{\tn,0}}{\AffStypMod{n,e}{\tn,0}}.
\end{equation}
Iterating this splicing, we arrive at a \emph{resolution} of the semitypical module $\AffStypMod{n,e}{\tn,0}$ in terms of Verma modules:
\begin{equation}
\res{\AffStypMod{n,e}{\tn,0}}{\AffVerMod{n-1/2,e}{\tn,0}}{\AffVerMod{n-3/2,e}{\tn,0}}{\AffVerMod{n-5/2,e}{\tn,0}}{\AffVerMod{n-7/2,e}{\tn,0}}.
\end{equation}
This is a (long) exact sequence, so we immediately obtain the character formula
\begin{equation}
\sch{\AffStypMod{n,e}{\tn,0}} = \sum_{m=0}^{\infty} \brac{-1}^m \sch{\AffVerMod{n-1/2-m,e}{\tn,0}}.
\end{equation}
Because spectral flow preserves module structure, it preserves exact sequences.  The generalisation of this result to the other semitypical irreducibles defined in \eqref{eqn:SFStyp} is therefore just
\begin{equation} \label{eqn:ResChStyp}
\sch{\sfmod{\ell}{\AffStypMod{n,e}{\tn,0}}} = \sum_{m=0}^{\infty} \brac{-1}^m \sch{\sfmod{\ell}{\AffVerMod{n-1/2-m,e}{\tn,0}}} = \sum_{m=0}^{\infty} \brac{-1}^m \sch{\AffVerMod{n-1/2 - 2 \ell - m,e + \ell k}{\tn,\ell \tk}}.
\end{equation}
Note that this sum converges in the sense that only finitely many terms in the sum will contribute to the multiplicity of any given weight space (where ``weight'' refers to the $N_0$- and $L_0$-eigenvalues).

It remains to analyse the atypical irreducibles in the same manner.  For this purpose, it is convenient to consider the submodule of $\AffVerMod{n-1,0}{\tn,0}$ generated by the $\tpsi^-_0$-descendant of the (generating) \hws{}.  This submodule is indecomposable, but reducible, and its character has the same form as that of the semitypical irreducibles, except that $e$ is set to $0$.  We may therefore denote it by $\AffStypMod{n-3/2,0}{\tn,0}$.  The quotient of the Verma module by this submodule is also indecomposable and reducible.  Indeed, it is isomorphic to $\AffStypMod{n-1/2,0}{\tn,0}$ according to the above discussion.  The relevance to the atypical irreducibles is that we now have the short exact sequence
\begin{equation}
\ses{\AffAtypMod{n-1,0}{\tn,0}}{\AffStypMod{n-1/2,0}{\tn,0}}{\AffAtypMod{n,0}{\tn,0}}.
\end{equation}
Splicing iteratively, we deduce resolutions for the atypical irreducibles, this time in terms of the semitypical indecomposables $\AffStypMod{n,0}{\tn,0}$.  The corresponding character formulae, generalised to all the atypicals defined by \eqref{eqn:SFAtyp}, is
\begin{align} \label{eqn:ResChAtyp}
\sch{\sfmod{\ell}{\AffAtypMod{n,0}{\tn,0}}} &= \sum_{m_1=0}^{\infty} \brac{-1}^{m_1} \sch{\sfmod{\ell}{\AffStypMod{n-1/2-m_1,0}{\tn,0}}} = \sum_{m_1,m_2=0}^{\infty} \brac{-1}^{m_1 + m_2} \sch{\sfmod{\ell}{\AffVerMod{n-1 - m_1 - m_2,0}{\tn,0}}} \notag \\
&= \sum_{m=0}^{\infty} \brac{-1}^m \brac{m+1} \sch{\sfmod{\ell}{\AffVerMod{n-1-m,0}{\tn,0}}} = \sum_{m=0}^{\infty} \brac{-1}^m \brac{m+1} \sch{\AffVerMod{n-1 - 2 \ell -m,\ell k}{\tn,\ell \tk}}.
\end{align}

It is now a simple matter to derive the S-transformations of the semitypical and atypical characters.  Since \thmref{thm:TypMod} applies to the indecomposable Verma module characters, we obtain
\begin{equation}
\begin{aligned}
\modS \set{\sch{\sfmod{\ell}{\AffStypMod{n,e}{\tn,0}}}} &= \sum_{m=0}^{\infty} \brac{-1}^m \underset{\RR^4}{\iiiint} \SMat{n-1/2 - 2 \ell - m,e + \ell k}{\tn,\ell \tk}{n',e'}{\tn',\te'} \sch{\AffTypMod{n',e'}{\tn',\te'}} \: \frac{\dd n' \: \dd e' \: \dd \tn' \: \dd \te'}{\tk^2}, \\
\modS \set{\sch{\sfmod{\ell}{\AffAtypMod{n,0}{\tn,0}}}} &= \sum_{m=0}^{\infty} \brac{-1}^m \brac{m+1} \underset{\RR^4}{\iiiint} \SMat{n-1 - 2 \ell - m,\ell k}{\tn,\ell \tk}{n',e'}{\tn',\te'} \sch{\AffTypMod{n',e'}{\tn',\te'}} \: \frac{\dd n' \: \dd e' \: \dd \tn' \: \dd \te'}{\tk^2}.
\end{aligned}
\end{equation}
Substituting \eqref{eqn:SMatTyp} and evaluating these sums, we arrive at the following result:

\begin{prop} \label{prop:NonTypMod}
The S-transformations of the semitypical $\DAKMSA{gl}{1}{1}$-supercharacters are given by
\begin{subequations}
\begin{equation}
\modS \set{\sch{\sfmod{\ell}{\AffStypMod{n,e}{\tn,0}}}} = \underset{\RR^4}{\iiiint} \SMatStyp{\ell}{n,e,\tn}{n',e'}{\tn',\te'} \sch{\AffTypMod{n',e'}{\tn',\te'}} \: \dd n' \: \dd \brac{e'/k} \: \dd \brac{k \tn' / \tk} \: \dd \brac{\te' / \tk},
\end{equation}
where the coefficients are given by
\begin{equation} \label{eqn:SMatStyp}
\SMatStyp{\ell}{n,e,\tn}{n',e'}{\tn',\te'} = \ee^{-2 \pi \ii \sqbrac{\brac{n \te' + \tn e' + \tn' e} / \tk - k \tn \te' / \tk^2}} \: \frac{\ee^{-2 \pi \ii n' \ell}}{2 \func{\cos}{\pi \te' / \tk}}.
\end{equation}
\end{subequations}
Similarly, the S-transformations of the atypical $\DAKMSA{gl}{1}{1}$-supercharacters are given by
\begin{subequations}
\begin{equation}
\modS \set{\sch{\sfmod{\ell}{\AffAtypMod{n,0}{\tn,0}}}} = \underset{\RR^4}{\iiiint} \SMatAtyp{\ell}{n,\tn}{n',e'}{\tn',\te'} \sch{\AffTypMod{n',e'}{\tn',\te'}} \: \dd n' \: \dd \brac{e'/k} \: \dd \brac{k \tn' / \tk} \: \dd \brac{\te' / \tk},
\end{equation}
with coefficients
\begin{equation} \label{eqn:SMatAtyp}
\SMatAtyp{\ell}{n,\tn}{n',e'}{\tn',\te'} = \ee^{-2 \pi \ii \sqbrac{\brac{n \te' + \tn e'} / \tk - k \tn \te' / \tk^2}} \: \frac{\ee^{-2 \pi \ii n' \ell}}{4 \func{\cos^2}{\pi \te' / \tk}}.
\end{equation}
\end{subequations}
\end{prop}

\noindent The overlines on the first entries of the S-matrix coefficients serve to remind us that the entry corresponds to a semitypical or atypical module.

\begin{cor} \label{cor:VacMod}
The S-transformation of the vacuum $\DAKMSA{gl}{1}{1}$-supercharacter is given by
\begin{subequations}
\begin{equation}
\modS \set{\sch{\AffAtypMod{0,0}{0,0}}} = \underset{\RR^4}{\iiiint} \SMatAtyp{0}{0,0}{n',e'}{\tn',\te'} \sch{\AffTypMod{n',e'}{\tn',\te'}} \: \dd n' \: \dd \brac{e'/k} \: \dd \brac{k \tn' / \tk} \: \dd \brac{\te' / \tk},
\end{equation}
with coefficients
\begin{equation} \label{eqn:SMatVac}
\SMatAtyp{0}{0,0}{n',e'}{\tn',\te'} = \frac{1}{4 \func{\cos^2}{\pi \te' / \tk}}.
\end{equation}
\end{subequations}
\end{cor}

Finally, we briefly consider the implications of these results for the modular invariance of the partition function.  In particular, \thmref{thm:TypMod} gives us the S-matrix with respect to a basis of $\DAKMSA{gl}{1}{1}$-supercharacters consisting of those of the typical and non-typical Verma modules.  It is easy to check that this S-matrix is symmetric and \emph{unitary}:
\begin{multline}
\underset{\RR^4}{\iiiint} \SMat{n,e}{\tn,\te}{n',e'}{\tn',\te'}^* \SMat{n,e}{\tn,\te}{n'',e''}{\tn'',\te''} \: \dd n \: \dd \brac{e/k} \: \dd \brac{k \tn / \tk} \: \dd \brac{\te / \tk} \\
= \func{\delta}{n' = n''} \func{\delta}{\frac{e'}{k} = \frac{e''}{k}} \func{\delta}{\frac{k \tn'}{\tk} = \frac{k \tn''}{\tk}} \func{\delta}{\frac{\te'}{\tk} = \frac{\te''}{\tk}}.
\end{multline}
It follows immediately that the diagonal partition function
\[
Z_{\text{diag.}} = \underset{\RR^4}{\iiiint} \abs{\sch{\AffTypMod{n,e}{\tn,\te}}}^2  \: \dd n \: \dd \brac{e/k} \: \dd \brac{k \tn / \tk} \: \dd \brac{\te / \tk}
\]
is modular invariant.  Similarly, unitarity and the symmetry of the S-matrix entries under conjugation $\brac{n,e,\tn,\te} \to \brac{-n,-e,-\tn,-\te}$ guarantee that the charge conjugation partition function is likewise invariant.  We remark that the supercharacters of the non-typical Verma modules could be further decomposed into those of their irreducible composition factors.  In particular, this makes it clear how the vacuum module appears in these partition functions.

\subsection{The Verlinde Formula}

Given the S-transformation formulae derived in \secref{sec:GL11Mod}, we can now apply a version of the Verlinde formula to derive the fusion coefficients.  Actually, because we are dealing with a logarithmic conformal field theory, meaning in particular that we expect the spectrum to include reducible but indecomposable modules, the Verlinde formula can only be expected to give the structure constants of the Grothendieck fusion ring.  This is the quotient of the true fusion ring in which one identifies each indecomposable module with the direct sum of its composition factors.  Alternatively, one identifies modules with the same character.  In this section, we assume that the Grothendieck ring is well-defined (that is, that fusion defines an exact functor from the spectrum to itself) and that its structure constants are given by the following continuum version of the Verlinde formula:
\begin{equation}
\fuscoeff{AB}{C} = \int_D \frac{\modS_{A,D} \modS_{B,D} \brac{\modS^{\dag}}_{C,D}}{\modS_{0,D}} \: \dd D.
\end{equation}
Here, $0$ refers to the vacuum module $\AffAtypMod{0,0}{0,0}$ and $D$ runs over a basis of the Grothendieck ring which we will take to consist of the (equivalence classes containing the) typical irreducible modules $\AffTypMod{n,e}{\tn,\te}$ with $\te \neq 0$ and the reducible Verma modules $\AffVerMod{n,e}{\tn,0} \equiv \AffTypMod{n,e}{\tn,0}$.  Because we want to interpret the coefficient $\fuscoeff{AB}{C}$ as describing the decomposition of the (Grothendieck) fusion of $A$ and $B$, we will always assume that $C$ is a basis element.  It then follows from the symmetry of the S-matrix in this basis (see \thmref{thm:TypMod}) that $\brac{\modS^{\dag}}_{C,D} = \modS_{C,D}^*$.

The easiest Verlinde coefficient to compute is that describing the Grothendieck fusion of an atypical and a typical.  Taking $A = \sfmod{\ell}{\AffAtypMod{n_1,0}{\tn_2,0}}$ and $B = \AffTypMod{n_2,e_2}{\tn_2,\te_2}$, the Verlinde formula becomes
\begin{equation} \label{eqn:VerAT}
\fuscoeff{\sfmod{\ell}{\AffAtypMod{n_1,0}{\tn_1,0}} \AffTypMod{n_2,e_2}{\tn_2,\te_2}}{\AffTypMod{n_3,e_3}{\tn_3,\te_3}} = \underset{\RR^4}{\iiiint} \frac{\SMatAtyp{\ell}{n_1, \tn_1}{n,e}{\tn,\te} \SMat{n_2,e_2}{\tn_2,\te_2}{n,e}{\tn,\te} \SMat{n_3,e_3}{\tn_3,\te_3}{n,e}{\tn,\te}^*}{\SMatAtyp{0}{0,0}{n,e}{\tn,\te}} \: \frac{\dd n \: \dd e \: \dd \tn \: \dd \te}{\tk^2}.
\end{equation}
Here, we should remark that the measure
\begin{equation}
\frac{\dd n \: \dd e \: \dd \tn \: \dd \te}{\tk^2} = \dd n \: \dd \brac{e/k} \: \dd \brac{k \tn / \tk} \: \dd \brac{\te / \tk}
\end{equation}
is the natural choice, given the rescaling properties \eqref{eqn:Rescaling} of the algebra.  Substituting \eqref{eqn:SMatTyp}, \eqref{eqn:SMatAtyp} and \eqref{eqn:SMatVac} into \eqref{eqn:VerAT}, we see that the denominator of the atypical S-matrix entry cancels that of the vacuum entry and the integrand becomes a pure phase.  In fact, the integral over $\RR^4$ separates into four integrals over $\RR$:
\begin{equation}
\begin{gathered}
\int_{\RR} \ee^{-2 \pi \ii \sqbrac{\brac{\te_2 - \te_3} / \tk + \ell} n} \: \dd n, \qquad
\int_{\RR} \ee^{-2 \pi \ii \sqbrac{\brac{e_2 - e_3} / k - \brac{\te_2 - \te_3} / \tk} k \tn / \tk} \: \dd \brac{k \tn / \tk}, \\
\int_{\RR} \ee^{-2 \pi \ii k \brac{\tn_1 + \tn_2 - \tn_3} / \tk \cdot e/k} \: \dd \brac{e/k}, \qquad
\int_{\RR} \ee^{-2 \pi \ii \sqbrac{n_1 + n_2 - n_3 + 2 \brac{\te_2 - \te_3} / \tk - k \brac{\tn_1 + \tn_2 - \tn_3} / \tk} \te / \tk} \: \dd \brac{\te / \tk}.
\end{gathered}
\end{equation}
These are easily evaluated and one arrives at
\begin{multline}
\fuscoeff{\sfmod{\ell}{\AffAtypMod{n_1,0}{\tn_1,0}} \AffTypMod{n_2,e_2}{\tn_2,\te_2}}{\AffTypMod{n_3,e_3}{\tn_3,\te_3}} \\
= \func{\delta}{n_3 = n_1 + n_2 - 2 \ell} \func{\delta}{\frac{e_3}{k} = \frac{e_2 + \ell k}{k}} \func{\delta}{\frac{k \tn_3}{\tk} = \frac{k \brac{\tn_1 + \tn_2}}{\tk}} \func{\delta}{\frac{\te_3}{\tk} = \frac{\te_2 + \ell \tk}{\tk}}.
\end{multline}
The corresponding Grothendieck fusion rule is therefore
\begin{align}
\sqbrac{\sfmod{\ell}{\AffAtypMod{n_1,0}{\tn_1,0}}} \fuse \sqbrac{\AffTypMod{n_2,e_2}{\tn_2,\te_2}} &= \underset{\RR^4}{\iiiint} \fuscoeff{\sfmod{\ell}{\AffAtypMod{n_1,0}{\tn_1,0}} \AffTypMod{n_2,e_2}{\tn_2,\te_2}}{\AffTypMod{n_3,e_3}{\tn_3,\te_3}} \sqbrac{\AffTypMod{n_3,e_3}{\tn_3,\te_3}} \: \frac{\dd n_3 \: \dd e_3 \: \dd \tn_3 \: \dd \te_3}{\tk^2} \nonumber \\
&= \sqbrac{\AffTypMod{n_1 + n_2 - 2 \ell , e_2 + \ell k}{\tn_1 + \tn_2 , \te_2 + \ell \tk}},
\end{align}
where the square brackets around a module indicate the equivalence class of the module in the Grothendieck ring.

The computations for (Grothendieck) fusing semitypicals with typicals, or typicals with typicals, are similar, using \eqref{eqn:SMatStyp} in the appropriate place instead of \eqref{eqn:SMatAtyp}.  The main difference is that the denominators appearing in the semitypical and vacuum S-matrix entries no longer cancel.  Instead, one obtains a pure phase multiplied by $2 \tfunc{\cos}{\pi \te' / \tk}$, or its square, hence the integrand may be written as a sum of two, or four, pure phases, respectively.  The actual integration proceeds as before.
%
We therefore turn to the Grothendieck fusion rules for non-typicals with non-typicals.  When fusing two semitypicals, the S-matrix entries' denominators cancel that coming from the vacuum in the Verlinde formula and the computations proceed as in the atypical by typical computation.  In contrast, the semitypical by atypical and atypical by atypical computations involve an integrand which is a pure phase divided by a non-trivial denominator.  Evaluation proceeds after re-expanding the denominator as follows:
\begin{equation}
\frac{1}{2 \func{\cos}{\pi \te / \tk}} = \sum_{m=0}^{\infty} \brac{-1}^m \ee^{\ii \pi \brac{2m+1} \te / \tk}, \qquad
\frac{1}{4 \func{\cos^2}{\pi \te / \tk}} = \sum_{m=0}^{\infty} \brac{-1}^m (m+1) \ee^{2 \pi \ii \brac{m+1} \te / \tk}.
\end{equation}
The Verlinde formula therefore gives the Grothendieck fusion coefficients as infinite sums of delta functions in these cases.  The corresponding Grothendieck fusion rules therefore decompose into infinite sums of typicals (actually, reducible Verma modules).  However, these infinite sums can be recognised as corresponding to semitypicals and atypicals using the resolution formulae \eqref{eqn:ResChStyp} and \eqref{eqn:ResChAtyp}.

We summarise the results of these computations:

\begin{prop} \label{prop:GrFusion}
Assuming that the Verlinde formula correctly gives the coefficients of the Grothendieck fusion ring, the product induced by the fusion product on the associated Grothendieck ring of $\DAKMSA{gl}{1}{1}$ is given, on the basis of (equivalence classes of) irreducibles by
\begin{equation}
\begin{gathered}
\sqbrac{\sfmod{\ell}{\AffAtypMod{n_1,0}{\tn_1,0}}} \fuse \sqbrac{\AffTypMod{n_2,e_2}{\tn_2,\te_2}} = \sqbrac{\AffTypMod{n_1 + n_2 - 2 \ell , e_2 + \ell k}{\tn_1 + \tn_2 , \te_2 + \ell \tk}},
\\
\sqbrac{\sfmod{\ell}{\AffStypMod{n_1,e_1}{\tn_1,0}}} \fuse \sqbrac{\AffTypMod{n_2,e_2}{\tn_2,\te_2}} = \sqbrac{\AffTypMod{n_1 + n_2 - 2 \ell + 1/2, e_1 + e_2 + \ell k}{\tn_1 + \tn_2 , \te_2 + \ell \tk}} \oplus \sqbrac{\AffTypMod{n_1 + n_2 - 2 \ell - 1/2, e_1 + e_2 + \ell k}{\tn_1 + \tn_2 , \te_2 + \ell \tk}},
\\
\sqbrac{\AffTypMod{n_1,e_1}{\tn_1,\te_1}} \fuse \sqbrac{\AffTypMod{n_2,e_2}{\tn_2,\te_2}} = \sqbrac{\AffTypMod{n_1 + n_2 + 1, e_1 + e_2}{\tn_1 + \tn_2 , \te_1 + \te_2}} \oplus 2 \sqbrac{\AffTypMod{n_1 + n_2, e_1 + e_2}{\tn_1 + \tn_2 , \te_1 + \te_2}} \oplus \sqbrac{\AffTypMod{n_1 + n_2 - 1, e_1 + e_2}{\tn_1 + \tn_2 , \te_1 + \te_2}},
\\
\sqbrac{\sfmod{\ell_1}{\AffStypMod{n_1,e_1}{\tn_1,0}}} \fuse \sqbrac{\sfmod{\ell_2}{\AffStypMod{n_2,e_2}{\tn_2,0}}} = \sqbrac{\AffTypMod{n_1 + n_2 - 2 \brac{\ell_1 + \ell_2}, e_1 + e_2 + \brac{\ell_1 + \ell_2} k}{\tn_1 + \tn_2 , \brac{\ell_1 + \ell_2} \tk}},
\\
\begin{aligned}
\sqbrac{\sfmod{\ell_1}{\AffAtypMod{n_1,0}{\tn_1,0}}} \fuse \sqbrac{\sfmod{\ell_2}{\AffStypMod{n_2,e_2}{\tn_2,0}}} &= \bigoplus_{m=0}^{\infty} \brac{-1}^m \sqbrac{\AffTypMod{n_1 + n_2 - 2 \brac{\ell_1 + \ell_2} - m - 1/2 , e_2 + \brac{\ell_1 + \ell_2} k}{\tn_1 + \tn_2 , \brac{\ell_1 + \ell_2} \tk}} \\
&= \sqbrac{\sfmod{\ell_1 + \ell_2}{\AffStypMod{n_1 + n_2 , e_2}{\tn_1 + \tn_2 , 0}}},
\end{aligned}
\\
\begin{aligned}
\sqbrac{\sfmod{\ell_1}{\AffAtypMod{n_1,0}{\tn_1,0}}} \fuse \sqbrac{\sfmod{\ell_2}{\AffAtypMod{n_2,0}{\tn_2,0}}} &= \bigoplus_{m=0}^{\infty} \brac{-1}^m \brac{m+1} \sqbrac{\AffTypMod{n_1 + n_2 - 2 \brac{\ell_1 + \ell_2} - m - 1 , \brac{\ell_1 + \ell_2} k}{\tn_1 + \tn_2 , \brac{\ell_1 + \ell_2} \tk}} \\
&= \sqbrac{\sfmod{\ell_1 + \ell_2}{\AffAtypMod{n_1 + n_2 , 0}{\tn_1 + \tn_2 , 0}}}.
\end{aligned}
\end{gathered}
\end{equation}
\end{prop}

\noindent As with \propref{prop:TPRules}, we have not fully decomposed these Grothendieck fusion rules into equivalence classes of irreducibles.  The typical fuse typical result will decompose into semitypicals or atypicals if $\brac{\te_1 + \te_2} / \tk \in \ZZ$ and the semitypical fuse semitypical result always decomposes in the same way.  As before, it is easy but cumbersome to write down the full decompositions given the above results.

\subsection{Fusion}

Because we assume that the Verlinde formula correctly gives the structure constants of the Grothendieck fusion ring, we consequently obtain a huge amount of information about the genuine fusion ring.  For example, when a Grothendieck fusion rule decomposes into a single equivalence class of an irreducible, then we may lift this decomposition to the corresponding fusion rule.  Similarly, if a Grothendieck decomposition turns out to be a direct sum of irreducible equivalence classes and no two of the corresponding irreducibles can be glued together to form an indecomposable, then we can be sure that the fusion rule also decomposes as a direct sum of irreducibles.  One easy test for deciding that two modules cannot be glued together indecomposably is to consider the conformal dimensions of their states modulo $\ZZ$ --- if the conformal dimensions have different fractional parts, then no gluing is possible.  With this in mind, \propref{prop:GrFusion} immediately gives:

\begin{thm} \label{thm:FusionSummary}
Assuming that the Verlinde formula correctly gives the coefficients of the Grothendieck fusion ring, we may deduce the following (genuine) fusion rules involving irreducibles:
\begin{equation} \label{eqn:KnownFR}
\begin{gathered}
\AffAtypMod{n_1,0}{\tn_1,0} \fuse \AffAtypMod{n_2,0}{\tn_2,0} = \AffAtypMod{n_1 + n_2 , 0}{\tn_1 + \tn_2 , 0}, \qquad
\AffAtypMod{n_1,0}{\tn_1,0} \fuse \AffStypMod{n_2,e_2}{\tn_2,0} = \AffStypMod{n_1 + n_2 , e_2}{\tn_1 + \tn_2 , 0}, \qquad
\AffAtypMod{n_1,0}{\tn_1,0} \fuse \AffTypMod{n_2,e_2}{\tn_2,\te_2} = \AffTypMod{n_1 + n_2 , e_2}{\tn_1 + \tn_2 , \te_2}. \\
\AffStypMod{n_1 + n_2 , e_2}{\tn_1 + \tn_2 , 0} \fuse \AffTypMod{n_2,e_2}{\tn_2,\te_2} = \AffTypMod{n_1 + n_2 + 1/2 , e_1 + e_2}{\tn_1 + \tn_2 , \te_2} \oplus \AffTypMod{n_1 + n_2 - 1/2 , e_1 + e_2}{\tn_1 + \tn_2 , \te_2}.
\end{gathered}
\end{equation}
Moreover, these fusion rules respect spectral flow in the sense that
\begin{equation} \label{eqn:SFAssumption}
\sfmod{\ell_1}{\mathcal{M}_1} \fuse \sfmod{\ell_2}{\mathcal{M}_2} = \sfmod{\ell_1 + \ell_2}{\mathcal{M}_1 \fuse \mathcal{M}_2}.
\end{equation}
\end{thm}

\noindent We remark that the semitypical by typical fusion follows because \eqref{eqn:GL11ConfDim} gives
\begin{equation}
\Delta^{n_1 + n_2 + 1/2 , e_1 + e_2}_{\tn_1 + \tn_2 , \te_2} - \Delta^{n_1 + n_2 - 1/2 , e_1 + e_2}_{\tn_1 + \tn_2 , \te_2} = \frac{\te_2}{\tk},
\end{equation}
which, by the typicality condition of \propref{prop:TypChar}, is not an integer.

These fusion rules should be compared with the tensor product rules computed for $\DSLSA{gl}{1}{1}$ in \eqnDref{TP:Ax}{TP:SxT} --- they are essentially the same.  To make this precise, recall from \secref{sec:AffGL11Reps} that when the $\tE_0$-eigenvalue is restricted to $-1 < \te / \tk < 1$, the non-trivial singular vectors may only occur at grade $0$ and then only when $\te = 0$.  For this range of $\te$, one therefore has a rather natural bijective correspondence between $\DAKMSA{gl}{1}{1}$-modules and the $\DSLSA{gl}{1}{1}$-modules formed by their zero-grade subspaces.  To put it differently, the induced module construction gives a bijection between $\DSLSA{gl}{1}{1}$-modules and $\DAKMSA{gl}{1}{1}$-modules when $\te / k$ is so restricted.  With this correspondence, we can formulate a plausible conjecture:
\begin{quote}
When $-\tfrac{1}{2} \leqslant \te_1 / \tk < \tfrac{1}{2}$ and $-\tfrac{1}{2} < \te_2 / \tk \leqslant \tfrac{1}{2}$, the fusion product of two $\DAKMSA{gl}{1}{1}$-modules with (respective) $\tE_0$-eigenvalues $\te_1$ and $\te_2$ is induced from the $\DSLSA{gl}{1}{1}$-module obtained from the tensor product of the $\DSLSA{gl}{1}{1}$-modules whose inductions are those being fused.  Fusion products involving modules with $\te_1$ and $\te_2$ outside the specified ranges may be computed from the above correspondence using spectral flow and \eqnref{eqn:SFAssumption}.
\end{quote}
Of course, \eqref{eqn:SFAssumption} is itself conjectural.  We have only verified that this relation holds for the fusion rules in \thmref{thm:FusionSummary} up to the assumption that the Verlinde formula is valid.

The first part of this conjecture, dealing with the fusion of modules with restricted $\tE_0$-eigenvalues, can actually be proven.  The argument is not difficult, but it relies on a reasonably detailed understanding of the celebrated Nahm-Gaberdiel-Kausch algorithm for fusion \cite{NahQua94,GabInd96}.  We will not repeat the argument here for the Takiff superalgebra $\DAKMSA{gl}{1}{1}$ --- it is practically identical to that given for the non-Takiff algebra $\AKMSA{gl}{1}{1}$ in \cite[Sec.~3.3]{CreRel11}.  Accepting this allows us to extend the fusion rules of \eqref{eqn:KnownFR} to include
\begin{equation} \label{FR:SxS}
\AffStypMod{n_1 , e_1}{\tn_1 , 0} \fuse \AffStypMod{n_2 , e_2}{\tn_2 , 0} = \AffStypMod{n_1 + n_2 + 1/2 , e_1 + e_2}{\tn_1 + \tn_2 , 0} \oplus \AffStypMod{n_1 + n_2 - 1/2 , e_1 + e_2}{\tn_1 + \tn_2 , 0},
\end{equation}
when $e_1 + e_2 \neq 0$.  This follows now from \eqnref{TP:SxS}, as does the fact that the result of this fusion when $e_1 + e_2 = 0$ is an indecomposable $\AffProjMod{n_1 + n_2 , 0}{\tn_1 + \tn_2 , 0}$ induced from the $\DSLSA{gl}{1}{1}$-module $\ProjMod{n_1 + n_2 , 0}{\tn_1 + \tn_2 , 0}$.  We can also now include the typical fusion rule
\begin{equation} \label{FR:TxT}
\AffTypMod{n_1 , e_1}{\tn_1 , \te_1} \fuse \AffTypMod{n_2 , e_2}{\tn_2 , \te_2} = \AffTypMod{n_1 + n_2 + 1 , e_1 + e_2}{\tn_1 + \tn_2 , \te_1 + \te_2} \oplus \: \AffGenTypMod{n_1 + n_2 , e_1 + e_2}{\tn_1 + \tn_2 , \te_1 + \te_2}{2} \oplus \AffTypMod{n_1 + n_2 - 1 , e_1 + e_2}{\tn_1 + \tn_2 , \te_1 + \te_2}
\end{equation}
which follows from \eqnref{TP:TxT} when $-\tfrac{1}{2} \leqslant \te_1 / \tk < \tfrac{1}{2}$, $-\tfrac{1}{2} < \te_2 / \tk \leqslant \tfrac{1}{2}$ and $\te_1 + \te_2 \neq 0$.  The module $\AffGenTypMod{n,e}{\tn,\te}{2}$ is that induced from the $\DSLSA{gl}{1}{1}$-module $\GenTypMod{n,e}{\tn,\te}{2}$, as usual.  We will not say anything more detailed about this fusion rule when $\te_1 + \te_2 = 0$ because we did not detail the result of the corresponding (rather complicated) tensor product decomposition in \secref{sec:TPFinGL11Reps}.  It is also worth noting that the fusion rules \eqref{FR:SxS} and \eqref{FR:TxT} may be extended to other $\te_1$ and $\te_2$ if we \emph{assume} that \eqref{eqn:SFAssumption} holds.

One thing that should be made explicit now is that the action of the Virasoro zero mode $L_0$ on the indecomposables $\AffGenTypMod{n,e}{\tn,\te}{2}$, with $\te \neq 0$, is almost always non-semisimple.  This follows because the action of $L_0$ on the ground states may be computed from the action of the $\DSLSA{gl}{1}{1}$ Casimirs $Q_1$ and $Q_2$ on the analogous $\DSLSA{gl}{1}{1}$-module $\GenTypMod{n,e}{\tn,\te}{2}$, given in \eqref{eqn:CasimirAction}.  If $\ket{v}$ denotes a ground state generator of $\AffGenTypMod{n,e}{\tn,\te}{2}$ with $\brac{\tN_0 - \tn} \ket{v} = \ket{w} \neq 0$, then we have
\begin{equation}
\brac{L_0 - \Delta^{n,e}_{\tn,\te}} \ket{v} = \frac{k}{\tk} \brac{\frac{e}{k} - \frac{\te}{\tk}} \ket{w}.
\end{equation}
We may therefore now conclude that \cfts{} based on the Takiff superalgebra $\DAKMSA{gl}{1}{1}$ are \emph{logarithmic}.

Finally, we note that an obvious corollary of \thmref{thm:FusionSummary} is that the atypicals $\sfmod{\ell}{\AffAtypMod{n,0}{\tn,0}}$ are all simple currents (invertible elements) of the fusion ring.  They can therefore be used to construct extended algebras, some of which may be of interest.  The extended algebras of $\AKMSA{gl}{1}{1}$ are known to include \cite{CreWAl11}:
\begin{itemize}
\item The direct sum of a free complex fermion and a set of $\beta \gamma$ ghosts.
\item The affine superalgebra $\AKMSA{sl}{2}{1}$ at levels $1$ and $-\tfrac{1}{2}$.
\item The Bershadsky-Polyakov algebra $W_3^{\brac{2}}$ at levels $0$ and $-\tfrac{5}{3}$.
\item The $\mathcal{N} = 2$ superconformal algebra of central charges $1$ and $-1$.
\item The Feigin-Semikhatov algebras $W_n^{\brac{2}}$ at various levels.
\end{itemize}
It therefore seems likely that the extended algebras of $\DAKMSA{gl}{1}{1}$ will include Takiff versions of these.  Moreover, if we choose the simple current so that its states have integer conformal dimensions, then one expects to be able to construct non-diagonal modular invariant partition functions that correspond to the diagonal partition function of the extended algebra.  We shall not consider extended algebras here, leaving their study for the future.

\section{Conclusions}

In this paper, we have considered a class of non-semisimple Lie algebras that we have referred to as Takiff algebras, from the point of view of representation theory and, in the affine case, conformal field theories generated by a Sugawara-type construction (\thmref{thm:ExtSugawara}).  The construction is universal for any affine Kac-Moody algebra or superalgebra.  We then concentrated on the Takiff superalgebra of $\SLSA{gl}{1}{1}$ and its affinisation as the simplest non-trivial examples; these are expected to demonstrate some of the basic features and structures typical for Takiff algebras of higher rank.  The classification of irreducible highest weight modules of the non-affine Takiff superalgebra $\ext{\SLSA{gl}{1}{1}}$ revealed many similarities to that of the non-Takiff superalgebra $\SLSA{gl}{1}{1}$, resulting in three classes of modules:  In addition to the generic (typical) class, we divided the remainder into those of atypicality rank $1$, which we called semitypical, and those of rank $2$, which were referred to as atypical.

By considering tensor products of the irreducible highest weight $\ext{\SLSA{gl}{1}{1}}$-modules, we were able to generate an infinite set of non-isomorphic indecomposable modules whose structures were partially unravelled.  In contrast to the non-Takiff case, in which the indecomposables can be completely classified and projectives/injectives identified, the zoo of indecomposables in the Takiff case seems to be far richer, though we do not claim any unclassifiability result.  Nevertheless, we succeeded in computing (\propref{prop:TPRules}) the product on the Grothendieck ring of $\ext{\SLSA{gl}{1}{1}}$-modules.

The main thrust of this paper was, however, related to the \cft{} structures associated with the affine Takiff algebra $\DAKMSA{gl}{1}{1}$.  The submodule structure of the Verma modules was completely elucidated using a spectral flow automorphism, leading to a complete classification of irreducible \hwms{}.  As with $\DSLSA{gl}{1}{1}$, these fell into three classes:  Typical, semitypical and atypical.  Character formulae followed immediately.  Our main result (\thmref{thm:TypMod}) is then the determination of the modular transformation properties of the supercharacters.  Here, we were able to generalise the usual affine S- and T-transformations so as to prove that the typical module supercharacters carry a representation of the modular group, albeit one of uncountably-infinite dimension.  The known structures of the semitypical and atypical Verma modules then allowed us to deduce the S-transformations of the supercharacters of their irreducible quotients.

Having at hand the explicit form of these S-transformations, we applied them to a formal calculation of fusion coefficients using a continuum version of the famous Verlinde formula.  This yielded non-negative integers (when properly interpreted) which gives strong evidence for the consistency of our results and, in particular, our thesis that these Takiff (super)algebras give rise to \cfts{}.  We then interpreted these Verlinde coefficients as the structure constants of the Grothendieck fusion ring and determined this ring explicitly (\propref{prop:GrFusion}).  We concluded by discussing the possibilities for lifting these results on the Grothendieck fusion ring to the genuine fusion ring of $\DAKMSA{gl}{1}{1}$ and the consistency of the fusion results with our tensor product computations for $\DSLSA{gl}{1}{1}$.  This led to the explicit confirmation that the \cft{} we propose, based on a chiral $\DAKMSA{gl}{1}{1}$ symmetry, is logarithmic in nature.

Summarizing, we see most of the typical features of logarithmic conformal field theory implemented
in the Takiff superalgebra theory considered here.  In particular, we have noted that there are modules on which the zero modes $\tN_0$ and $L_0$ act non-semisimply and that fusion generates examples of indecomposable modules from irreducible ones.  One interesting feature of this theory is the non-semisimple action of the affine Cartan element $\tN_0$ on the typical irreducibles $\AffTypMod{n,e}{\tn,\te}$ and their indecomposable cousins $\AffGenTypMod{n,e}{\tn,\te}{m}$.  Aside from the (logarithmic) free boson, we are not aware of any other theory exhibiting this.  The effect on correlation functions is, however, masked by the fact that $L_0$ also acts non-semisimply on these modules.  There are therefore two-point functions with logarithmic singularities as one requires in a \lcft{}.  To isolate the effect of the non-semisimplicity of $\tN_0$, one can instead consider the Verma modules.  Here, one sees that the effect is minimal:  As usual, the eigenvectors in any non-trivial Jordan cell for $\tN_0$ end up being null and the non-trivial coupling of the two-point function occurs between the eigenvectors and their Jordan partners.  However, no logarithms are encountered in the corresponding correlators.

Of course, we have only considered here the simplest example of an affine Takiff (super)algebra.  The next obvious candidate for detailed study is affine Takiff $\SLA{sl}{2}$, which we might denote by $\DAKMA{sl}{2}$.  The situation here is significantly more complicated, because we expect that the full spectrum of irreducibles will involve modules which are not highest weight with respect to any choice of triangular decomposition.  Unfortunately, the mathematical properties of such modules seem to be completely unknown for Takiff superalgebras and we hope to report on this and \cfts{} based on $\DAKMA{sl}{2}$ in the future \cite{BR}.

\section*{Acknowledgements}

AB would like to thank Maria Gorelik for interesting discussions and the Einstein center of Weizmann Institute for support. He thanks also the University of York for hospitality where this work was finished, and support from UK EPSRC grant number EP/H000054/1.  DR's research is supported by the Australian Research Council Discovery Project DP1093910.  We would also like to thank Thomas Quella, Volker Schomerus and Jan Troost for discussions related to the work reported here and gratefully acknowledge the Institut Henri Poincar\'{e} and, in particular, the organisers of the recent trimestre ``Advanced Conformal Field Theory and Applications'', for hospitality during a phase of this project.

\raggedright


\end{document}